\documentclass[reprint,aps,prl,nofootinbib,superscriptaddress]{revtex4-2}
\usepackage[utf8]{inputenc}
\usepackage{amsmath}
\usepackage{amsfonts}
\usepackage{amsthm}
\usepackage{amssymb}
\usepackage{graphicx}
\usepackage[usenames,dvipsnames]{color}
\usepackage{hyperref}
\usepackage{mathrsfs}
\usepackage{orcidlink}

\theoremstyle{definition}

\theoremstyle{remark}

\newcommand{\set}[1]{\left\{#1\right\}}

\def\actaa{\ref@jnl{Acta Astron.}}      
\usepackage[normalem]{ulem}

\usepackage[usenames]{xcolor}

\begin{document}

\title{Second-Generation Mass Peak in the Gravitational-Wave Population as a Probe of Globular Clusters}

\author{Yonadav Barry Ginat\,\orcidlink{0000-0003-1992-1910}}%
\email{yb.ginat@physics.ox.ac.uk}%
\affiliation{Rudolf Peierls Centre for Theoretical Physics, University of Oxford, Parks Road, Oxford, OX1 3PU, United Kingdom}%
\affiliation{New College, Holywell Street, Oxford, OX1 3BN, United Kingdom}%
\author{Fabio Antonini}
\email{antoninif@cardiff.ac.uk}
\affiliation{Gravity Exploration Institute, School of Physics and Astronomy, Cardiff University, 5 The Parade, Cardiff, CF24 3AA, United Kingdom}
\author{Elizabeth Flanagan}
\affiliation{Gravity Exploration Institute, School of Physics and Astronomy, Cardiff University, 5 The Parade, Cardiff, CF24 3AA, United Kingdom}
\author{Mark Gieles\orcidlink{0000-0002-9716-1868}}
\affiliation{ICREA, Pg.~Llu\'is Companys 23, 08010 Barcelona, Spain}
\affiliation{Institut de Ciències del Cosmos (ICCUB), Universitat de Barcelona (UB), c.~Martí i Franqués, 1, 08028 Barcelona, Spain}
\affiliation{Institut d'Estudis Espacials de Catalunya (IEEC), Edifici RDIT, Campus UPC, 08860 Castelldefels (Barcelona), Spain}

\begin{abstract}
    Gravitational-wave observations have revealed an excess of binary black hole mergers with primary masses near $\sim 35\,M_\odot$. We show that if this feature originates from dynamical formation in dense stellar systems, and if the pair-instability supernova truncates the first-generation black hole mass spectrum, then second-generation mergers inevitably produce a second peak near $\sim 70\,M_\odot$. This structure reflects the suppression of first-generation black holes above a characteristic mass and the accumulation of merger remnants near twice that scale. Its location is robust, whereas its amplitude depends strongly on cluster initial conditions. Using a large suite of cluster population-synthesis models, we show that current gravitational-wave data already constrain the birth properties of globular clusters, irrespective of their overall contribution to the observed population. If clusters dominate mergers above the pair-instability scale, these constraints tighten further and imply a minimum first-generation merger rate of $\mathcal{R}(m_1 \leq 50\,M_\odot) \geq 0.099\,\mathrm{Gpc}^{-3}\,\mathrm{yr}^{-1}$ ($99\%$ confidence). We further show that a drop or gap in the secondary black hole mass spectrum is not a robust signature of a cluster origin for high-mass mergers within the pair-instability mass gap. A confirmed excess near $\sim 70\,M_\odot$ would support a dynamical origin of the $\sim 35\,M_\odot$ feature and provide independent evidence for a pair-instability mass gap with a lower edge at $\lesssim 50M_\odot$.
\end{abstract}

\maketitle

\paragraph{Introduction.}

The rapidly growing catalogue of gravitational-wave (GW) detections has opened a new window on the population of merging black holes~\cite{LIGOVirgo2016,LVK_GWTC3_2023,LVK2025GWTC4,LVK_GWTC4_2025_zenodo}. An important feature emerging from these analyses is an excess of mergers with a primary-mass near $\sim 35\,M_\odot$, either a peak or a knee, which has been  identified across multiple inference studies~\cite{2021ApJ...913L..19T,LVK_GWTC3_2023,LVK2025GWTC4,2025arXiv250915646B,Farahetal2024}. 
Understanding this feature is central to identifying the astrophysical environments that produce coalescing black holes.

Models of isolated binary evolution are able to explain this peak \cite{2019ApJ...882..121S, 2023MNRAS.520.5724B,Kishoreetal2025,2025arXiv250915646B}, but it is often ascribed to the dynamical formation of binaries in dense stellar systems, like globular clusters~\cite{Antoninietal2023}. In these environments, repeated few-body encounters harden black-hole binaries \cite{Heggie1975, HutBahcall1983} until GW emission drives them to coalesce~\cite{Monaghan1976a,Monaghan1976b,AnosovaOrlov1986,Mikkola1994,ValtonenKarttunen2006,Samsingetal2017,Samsing2018,StoneLeigh2019,GinatPerets2021a,Kol2021,Tranietal2024b,MarinGieles2024,Randoetal2025,Marinetal2025}. Mass segregation concentrates the most massive black holes in cluster cores, naturally favouring mergers among the heaviest available objects~\cite[e.g.][]{HeggieHut2003,BreenHeggie2013,Vaclacetal2018,Weatherfordetal2020,Arcaseddaetal2023a,Ishchenkoetal2024}. In this picture, the $\sim 35\,M_\odot$ peak arises from the interplay between stellar evolution and dynamical processing: the low metallicity of clusters produces an extended initial black-hole mass function, while dynamical evolution operates more strongly at higher masses, thereby biasing high the mass distribution~\cite{Antoninietal2023}. Consistent with this interpretation, analyses of the GW population indicate that binaries in this peak preferentially have similar masses and isotropic spin orientations, supporting a dynamical origin~\cite{Sridharetal2025,2025arXiv250915646B}.

A key process shaping the black-hole mass spectrum is pair-instability supernov{\ae} (PISN), which suppress the formation of black holes above a cut-off mass $m_\ast\sim 40$--$70\,M_\odot$~\cite[e.g.][]{FowlerHoyle1964,RakavyShaviv1967,Barkatetal1967,1968Ap&SS...2...96F,Fryer_2001,Woosley2016,Woosley2021}; this sets the characteristic mass scale of first-generation (1G) black holes~\cite{2021MNRAS.502L..40F,2020ApJ...890..113B,2017MNRAS.470.4739S,2020ApJ...902L..36F,2023MNRAS.526.4130H,2019ApJ...887...53F,2019ApJ...887...72L}. Yet, no clear mass gap is seen in the black-hole distribution inferred from GW data~\cite{2019ApJ...882L..24A,2021ApJ...913L...7A,2021ApJ...913L..23E,LVK_GWTC3_2023,ray_nonparametric_2023,2024PhRvX..14b1005C,2025PhRvD.112b3531A,LVK_GWTC1_2019,Abbott:2020gyp,LVK_GWTC_2,LVK2025GWTC4}. Dense stellar systems offer a way out of this discrepancy, by providing a natural way to populate this mass range through hierarchical mergers, where second-generation (2G) black holes (the merger remnants of 1G black holes) are retained by the cluster and merge again~\cite[e.g.][]{2019PhRvD.100d3027R,Rodriguezetal2022,BarberAntonini2025,Maietal2026,Ye_etal_2026,2019MNRAS.486.5008A,2023PhRvD.108h3012K}.

In this paper we use a large suite of cluster simulations to reveal a generic feature: if the observed peak near $\sim 35\,M_\odot$ has a dynamical origin in globular clusters, then a second peak near $\sim 70\,M_\odot$ should be present as the result of hierarchical mergers. This feature arises because merger remnants accumulate near twice the characteristic 1G mass scale, while the PISN process suppresses the formation of black holes in the same mass range. The result is a two-peak structure in the high-mass end of the mass distribution.

Current GW data already show tentative support for an excess merger rate near $m_1\sim 60$--$70\,M_\odot$~\cite{HernandezPalmese2025,HernandezPalmese2025b,Sridharetal2025,2026arXiv260306566A,PierraPapadopoulos2026}, although the statistical significance is limited, so far. 
While black holes above $50\,M_\odot$ are allowed by the uncertainties in nuclear reaction rates in massive stars \cite{Farmer2020} and stellar winds \cite{2021MNRAS.504..146V}, recent work suggests that mergers in this mass range may be hierarchical in origin, naturally explaining their large masses~\cite{2021ApJ...913L..19T,2025arXiv250819208M,2025arXiv251025579T,2025arXiv250819208M} and spin properties~\cite[e.g.,][]{2022ApJ...941L..39W,2023arXiv230302973L,Antoninietal2025a,Antoninietal2025b,Antoninietal2025c,Pierraetal2024}.

What has been missing, however, is evidence that cluster dynamics can generate enough hierarchical mergers to account for an excess near this mass at the observed rate. It has also remained unclear why such events should accumulate at 
this particular mass scale, and how they are connected to the broader black-hole mass spectrum at lower masses. Our results provide this connection, and show that the same feature can be used to constrain the initial conditions of globular clusters, via a direct comparison between models and  data.

The next observing runs will enlarge the catalogue of binary black-hole mergers substantially, and thus provide an immediate opportunity to test whether the tentative excess near $70\,M_\odot$ is real, and whether it is connected to the peak at $35\,M_\odot$. If confirmed, this structure would provide a new probe of dynamical black-hole assembly, and its existence will offer independent evidence for a PISN mass gap.

\paragraph{Cluster modelling and data inference.}
We model dynamically formed black-hole binaries in globular clusters using the fast cluster population-synthesis code \texttt{cBHBd}~\cite{AntoniniGieles2020a}, which reproduces merger rates consistent with Monte Carlo and $N$-body simulations~\cite{AntoniniGieles2020a,BarberAntonini2025}. The code evolves binaries statistically under repeated binary--single encounters and includes a realistic metallicity-dependent black-hole mass function, stellar-evolution mass loss, mass loss to gravitational waves (using the prescription of ref.~\cite[][Eq.~18]{Barause_etal2012}), natal kicks, dynamical ejections, gravitational-wave recoil, and black-hole growth through repeated mergers. In all models we adopt the PISN prescription of Ref.~\cite{2015MNRAS.451.4086S}, which yields a maximum 1G black-hole mass of $m_\ast\simeq 50\,M_\odot$.

We compute the differential merger rate from the catalogue of coalescence events produced by \texttt{cBHBd}:
\begin{equation}
    \frac{\mathrm{d}\mathcal{R}}{\mathrm{d}m_1}
    =
    K\rho_{\rm GC}
    \frac{\sum_{\rm cl} w_{\rm cl}\,\partial \dot{N}_{\rm mg}/\partial m_1}
    {\sum_{\rm cl} w_{\rm cl}\,M_{\rm cl,0}},
    \label{eqn:rate_definition}
\end{equation}
where $\dot{N}_{\rm mg}$ is the merger rate per cluster and
$w_{\rm cl}\equiv \phi_Z(Z|t_{\rm f})\phi_t(t_{\rm f})\phi_{M,0}(M_{\rm cl,0})M_{\rm cl,0}$
are weights set by the distributions of initial masses $M_{\rm cl,0}$, metallicities $Z$, and formation times $t_{\rm f}$ of the clusters. Above, the factor
\begin{equation}
    K \equiv \frac{\int M_{\rm cl,0}\,\phi_{M,0}(M_{\rm cl,0})\,\mathrm{d}M_{\rm cl,0}}{\int M\,\phi_{M}(M)\,\mathrm{d}M}
\end{equation}
is the ratio of the total mass of clusters in the initial conditions and at present~\cite{AntoniniGieles2020b}. Details of the distributions $\phi$ are given in the Supplemental Material (SM).

The cluster population is characterised by the parameters
$\{M_{\rm c},\Delta,\rho_{\rm h,i},\rho_{\rm GC}\}$.
Here, $M_{\rm c}$ is the exponential high-mass truncation of the initial cluster mass function,
$\phi_{\rm cl,0}\propto M_{\rm cl,0}^{-2}\exp[-M_{\rm cl,0}/(2M_{\rm c})]$,\footnote{The `2' in the denominator arises because we assume that clusters lose half of their initial mass by stellar evolution \cite[cf.][\S B]{AntoniniGieles2020b}.}
and $\Delta$ is the mass lost by a cluster over its life-time, assumed to be constant across clusters as in ref.~\cite{AntoniniGieles2020b}. These parameters specify the shape of both the initial cluster mass function and the present-day mass function as the result of dynamical evolution~\cite{Jordanetal2007}. 
The parameter $\rho_{\rm GC}$ is the mass density of globular clusters in the Universe today, such that the product $\rho_{\rm GC,0} \equiv K\rho_{\rm GC}$ is the initial mass density of globular clusters---that is, before mass loss by stellar evolution and tidal evaporation. Finally, 
$\rho_{\rm h,i}$ is the average mass density of clusters within their half-mass radii. 
We consider models where all clusters share the same initial half-mass density $\rho_{\rm h,i}$, as well as models with a mass-dependent density, $\rho_{\rm h,i}\propto M_{\rm cl}^\beta$ (see SM). The total number of clusters simulated is $\sim 4\times 10^7$. We emphasise that the models are not tuned to GW data: all simulations adopt the same set-up as ref.~\cite{Antoninietal2023}, except in-so-far as the initial black-hole spins were selected: here we consider models where the initial black holes had zero spins (consistent with theoretical models of BH formation \cite{Fuller2019}), and in the SM we discuss ones where the initial spins are drawn from a truncated Gaussian distribution centred at $\chi = 0.1$ with a width $0.3$, to align with the observed distribution in GWTC-4 \cite[][figs.~7, 17]{LVK2025GWTC4}.

We compare these models to the black-hole mass distribution we infer from the binary black-hole merger candidates in GWTC-4~\cite{LVK2025GWTC4}. The underlying distributions of primary and secondary masses are reconstructed non-parametrically, using Gaussian-process priors for both, while accounting for selection effects, as described in the SM.

\begin{figure}
    \centering
    \includegraphics[width=0.49\textwidth]{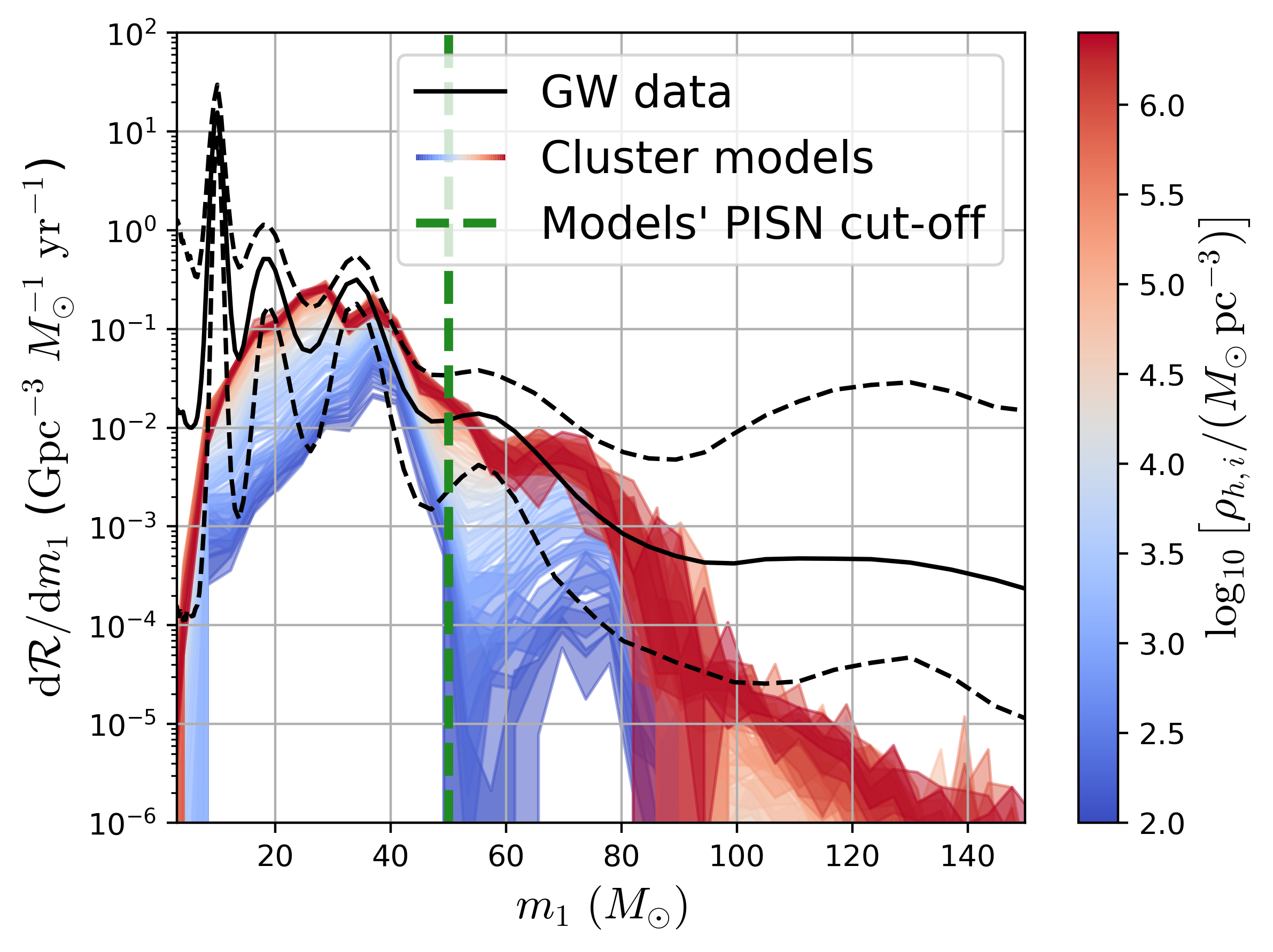}
    \includegraphics[width=0.49\textwidth]{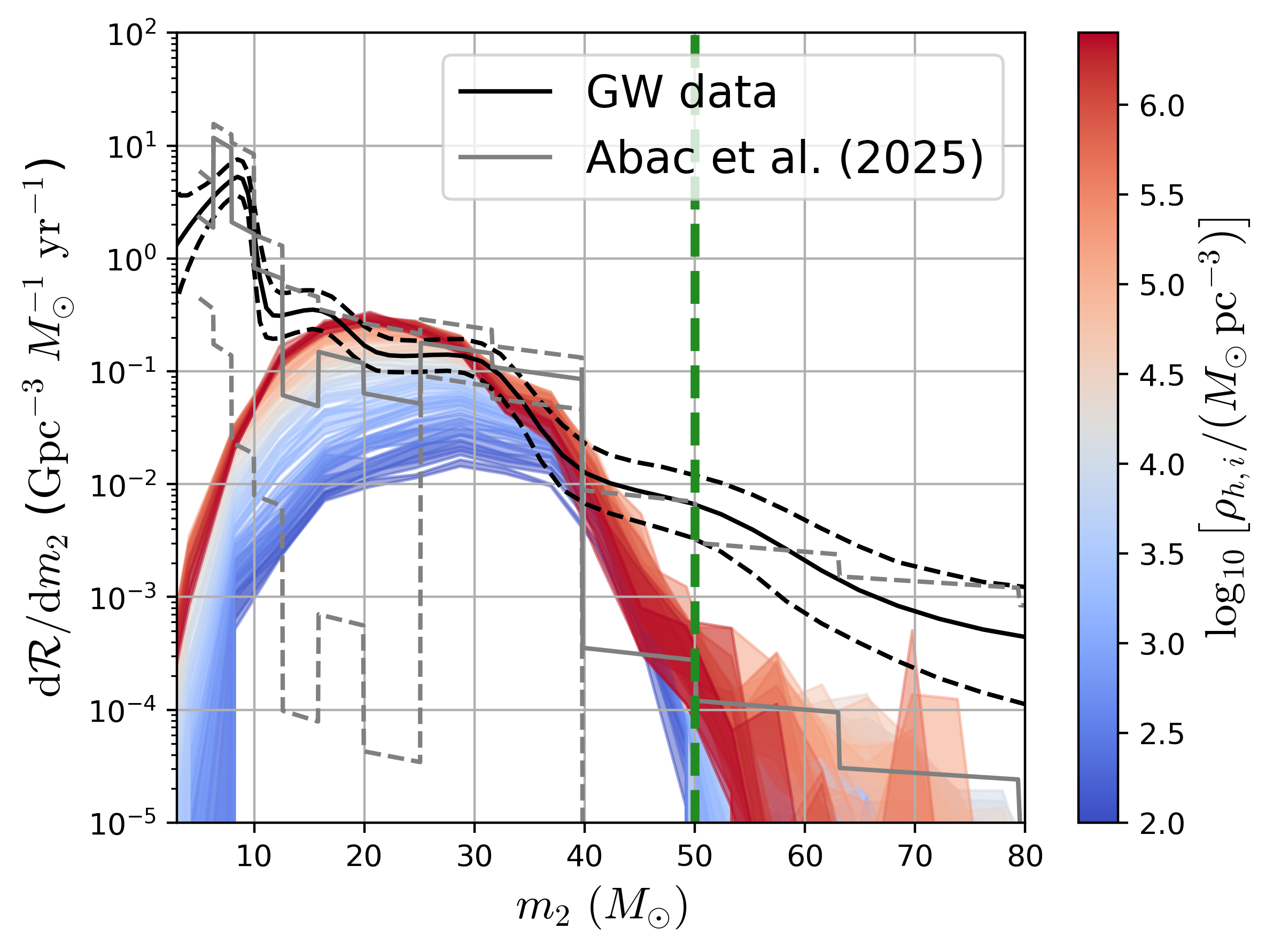}
    \caption{\emph{Top}: Local merger rate $\mathrm{d}\mathcal{R}/\mathrm{d}m_1$ for representative cluster models. We fix the cluster mass function and $\rho_{\rm GC}$ and vary the initial half-mass density $\rho_{\rm h,i}$. The LVK-inferred rate is shown for comparison (black). \emph{Bottom}: Corresponding secondary-mass distribution. For comparison, LVK's {\tt BGP} distribution $p(m_2)$~\cite{LVK2025GWTC4}, normalised to the total rate of our model, is shown in grey to illustrate the model dependence of the inferred distribution at high masses. Representative parameters are 
        $\Delta=10^{5.15}\,M_\odot$, $M_{\rm c}=10^{5.96}\,M_\odot$, and $\rho_{\rm GC}=4\times 10^{14}\,M_\odot\,\mathrm{Gpc}^{-3}$.
    }
    \label{fig:example_rates}
\end{figure}

\begin{figure*}
    \centering
    \includegraphics[width=0.5\textwidth]{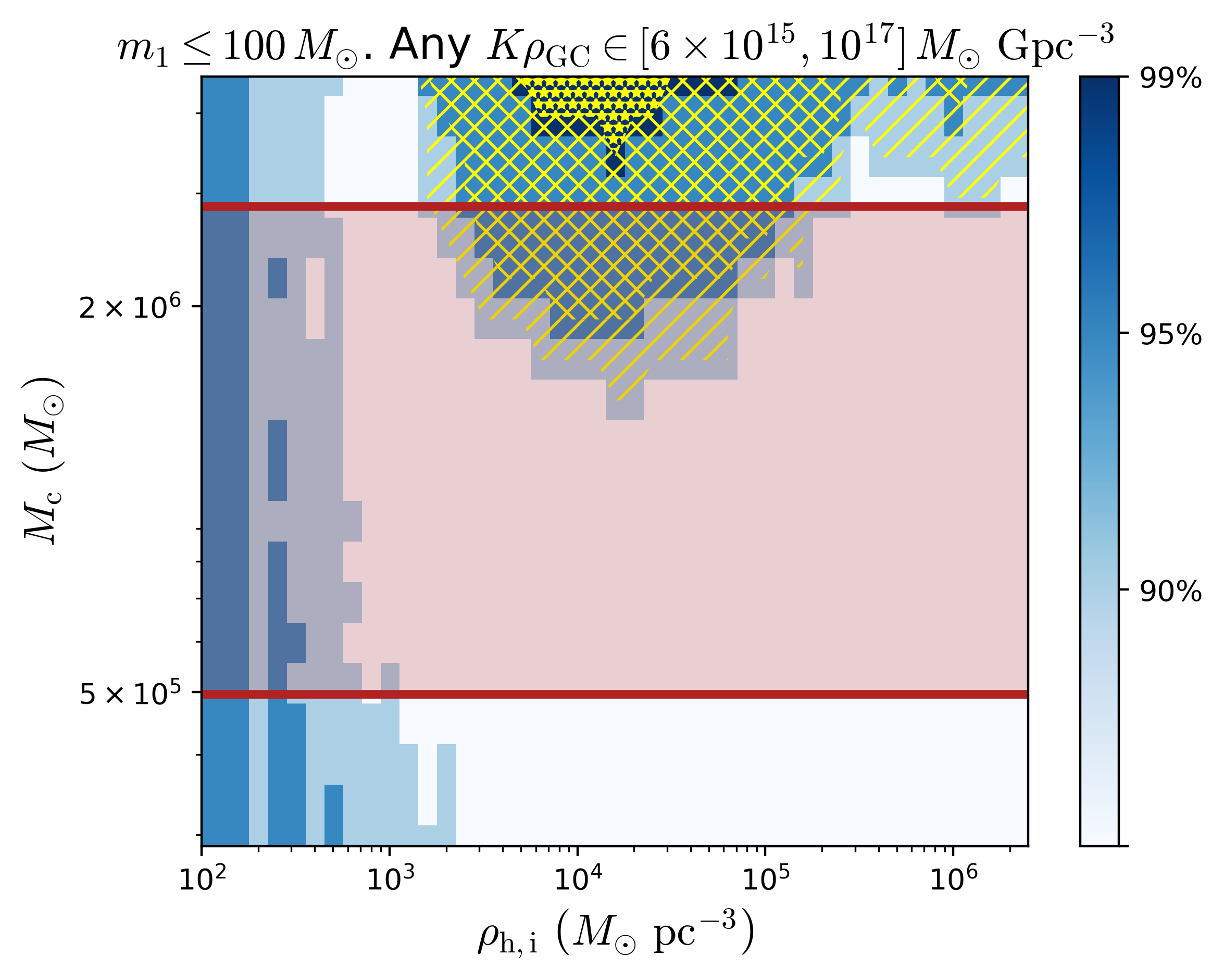}
    \includegraphics[width=0.48\textwidth]{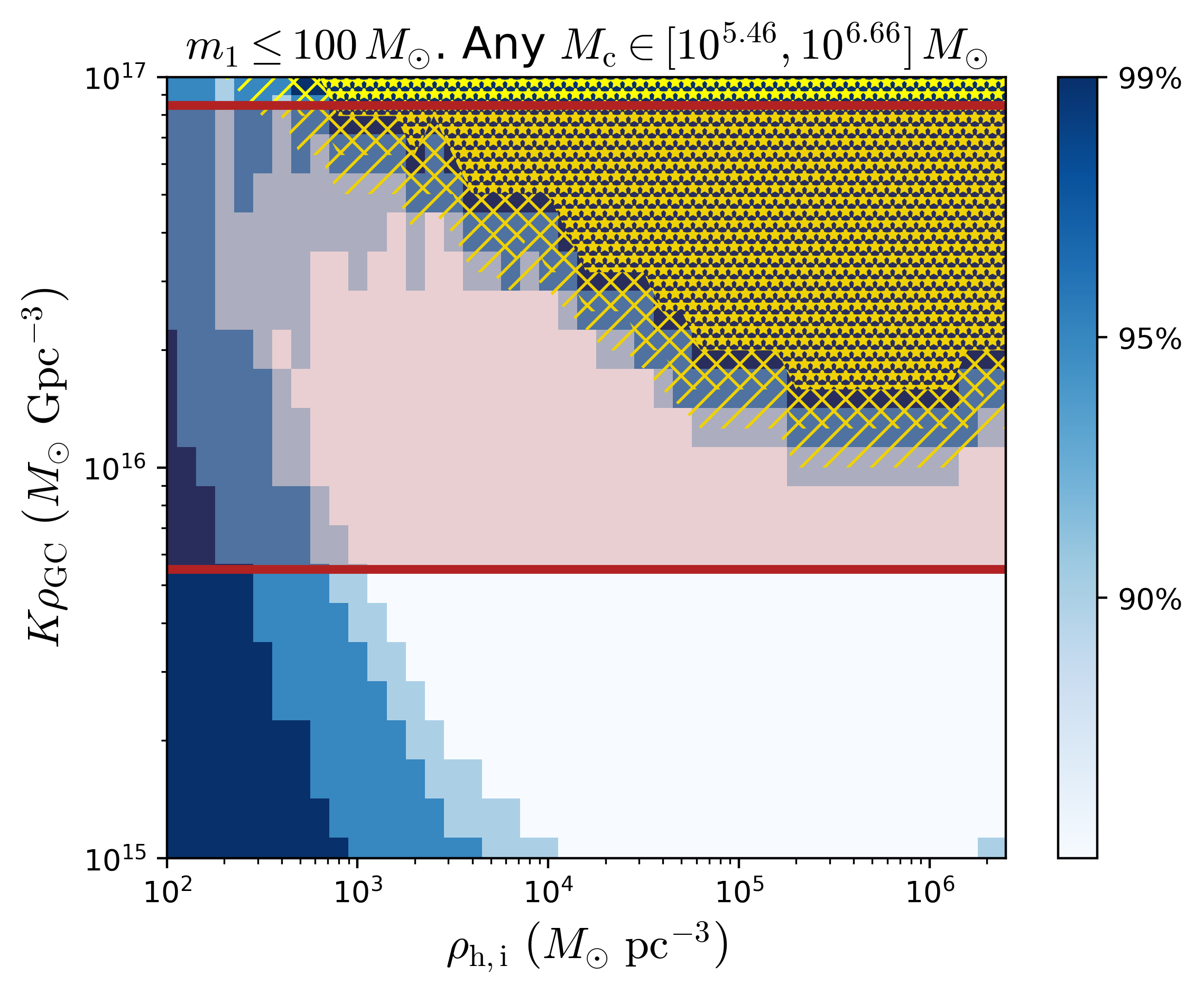}
    \caption{
        Cluster initial conditions constrained by GW data. Yellow hatching shows regions in which the cluster rate exceeds the LVK upper limit (stripes: $90\%$; crosses: $95\%$; asterisks: $99\%$). Coloured regions show stronger constraints obtained by additionally requiring that $\mathcal{R}(m_1\geq 50\,M_\odot)$ agree with the LVK-inferred high-mass rate. \emph{Left}: a point in $(M_{\rm c},\rho_{\rm h,i})$ is coloured if all models with $6\times10^{15}\,M_\odot\,\mathrm{Gpc}^{-3}\leq K\rho_{\rm GC}\leq 10^{17}\,M_\odot\,\mathrm{Gpc}^{-3}$ are in tension with the data. \emph{Right}: a point in $(K\rho_{\rm GC},\rho_{\rm h,i})$ is coloured if all models with $M_{\rm c}\in[10^{5.46},10^{6.66}]\,M_\odot$ are in tension. Red lines indicate the $95\%$ intervals for $M_{\rm c}$ and $K\rho_{\rm GC}$ inferred from globular-cluster observations and cosmological modelling~\cite{Jordanetal2007,AntoniniGieles2020b}.}
    \label{fig:constraints summary 1}
\end{figure*}

\paragraph{A second high-mass peak?}
Figure~\ref{fig:example_rates} shows the coalescence rate as a function of the primary and secondary masses, as predicted by \texttt{cBHBd}, for representative cluster-population parameters $\{M_{\rm c},\Delta,\rho_{\rm GC}\}$, with the initial half-mass density $\rho_{\rm h,i}$ varied.
Across the explored parameter space, the models produce two features in the primary-mass distribution consistently: a peak at $m_1\simeq 35\,M_\odot$ and a secondary peak near $m_1\simeq 70\,M_\odot$, followed by a rapid decline at higher masses. The relative height of the two peaks depends on the efficiency of hierarchical mergers, and therefore varies with cluster density and mass-function parameters, but their locations remain nearly fixed. Over the range
$30\,M_\odot\lesssim m_1\lesssim 100\,M_\odot$,
models with initial densities
$\rho_{\rm h,i}\gtrsim 10^4\,M_\odot\,\mathrm{pc}^{-3}$
lie within the $90$th percentile of the LVK-inferred rate for zero initial spins, while they do so for $\rho_{\rm h,i}\gtrsim 10^5\,M_\odot\,\mathrm{pc}^{-3}$ for LVK initial spins (see SM). This indicates that, for realistic initial conditions \cite{2024Natur.632..513A}, globular clusters can explain most, if not all, mergers in this mass range. 
At lower densities, the rate between the two peaks becomes progressively more suppressed, leaving a mass distribution with two prominent peaks and almost no mergers outside them.

In the SM, we show that a feature in the 1G merger distribution with centroid $m_0$ and width $\sigma_{\rm 1G}$ is transferred into a second-generation feature with centroid
$M_0\simeq \eta(\bar q)(1+\bar q)m_0$
and width
$\sigma_{\rm 2G}\simeq \eta(\bar q)(1+\bar q)\sigma_{\rm 1G}$,
where $\eta$ is the remnant mass fraction after GW emission and $\bar q$ is the characteristic mass ratio of the 1G$+$1G mergers. Because cluster dynamics favour near-equal-mass mergers, $\bar q\sim 1$ and $\eta\simeq 0.95$, a feature near $35\,M_\odot$ is  mapped to a wider feature at $\sim 65$--$70\,M_\odot$. 
The 2G remnant population therefore accumulates into a distinct second peak rather than a smooth high-mass tail. The location of the peak is  set mainly by the 1G mass scale and the PISN cut-off, and is largely insensitive to the details of cluster evolution. Its amplitude, however, depends strongly on cluster initial conditions since it reflects the efficiency
with which merger remnants are retained by the clusters
and take part into subsequent mergers. This is borne out by Fig.~\ref{fig:example_rates}, where models with different initial densities produce peaks at nearly the same mass but with different relative heights.

In the SM we report results for models with $\rho_{\rm h,i}\propto M_{\rm cl}^\beta$ and models with with non-zero initial spins. The same qualitative features are present there as well. Hence, the two-peak structure is not an artefact of assuming either a mass-independent initial cluster density or zero initial spins.

Most models also produce a secondary-mass distribution that lies within the inferred $90\%$ confidence band in the range
$20\,M_\odot\lesssim m_2\lesssim 40\,M_\odot$
[Fig.~\ref{fig:example_rates}, bottom panel]. One might have expected the secondary-mass distribution to drop sharply at the lower edge of the PISN gap because $2\mathrm{G}+2\mathrm{G}$ mergers are rare~\cite{HuiMaya}. However, this expectation has not previously been tested in a cluster population framework. We find no clear cut-off at the pair-instability scale ($m_*\simeq 50\,M_\odot$ in our models). The dominant $1\mathrm{G}+1\mathrm{G}$ merger rate declines steeply above $m_2\sim 30\,M_\odot$ and is strongly suppressed at the PISN edge, after which the distribution is refilled---and in many models even rises again near $50\,M_\odot$---predominantly due to $2\mathrm{G}+2\mathrm{G}$ mergers. This behaviour is consistent with the interpretation that the $70\,M_\odot$ feature in the primary-mass distribution arises mainly from $2\mathrm{G}+1\mathrm{G}$ mergers, while rare $2\mathrm{G}+2\mathrm{G}$ ones shape the secondary-mass distribution within the PISN gap. 
The absence of a sharp cut-off in the data therefore cannot be used to exclude a pure cluster origin for high-mass mergers directly~\cite[e.g.,][]{RayKalogera2026}. 
However, it also suggests that a gap in the secondary-mass distribution, if present, cannot be unambiguously associated with the onset of PISN~\cite{HuiMaya}.

At the highest masses, $m_1\gtrsim 100\,M_\odot$ and $m_2\gtrsim 40\,M_\odot$, the cluster channel predicts rates that are lower than those we  derived from the data. We do not draw strong conclusions from this comparison, however, since the inference in this regime is sensitive to small-number statistics and to assumptions on the priors. As illustrated in the bottom panel of figure~\ref{fig:example_rates}, for example, Ref.~\cite{LVK2025GWTC4} show that data are consistent with a secondary-mass distribution that plummets above $\simeq 40M_\odot$.
If future data end up confirming a persistent excess there, that would point to additional formation channels at the highest masses.

\paragraph{Cluster initial conditions from GWs.}
The relative height of the two peaks depends sensitively on the cluster parameters
$\{M_{\rm c},\rho_{\rm GC},\rho_{\rm h,i},\Delta\}$,
and can therefore be used to constrain globular-cluster formation and evolution. 

We first derive constraints that make no assumption about the contribution of clusters to the merger rate. To do so, we exclude regions of parameter space whose predicted rates exceed the LVK-inferred rate at any
$m_1\leq 100\,M_\odot$. This criterion is channel-agnostic: any additional formation channel can only increase the total rate and therefore cannot rescue a cluster model that already over-predicts the data.

Figure~\ref{fig:constraints summary 1} illustrates this approach. We compute the coalescence rate over a grid in
$\{M_{\rm c},\rho_{\rm h,i},K\rho_{\rm GC}\}$
and identify models that exceed the LVK-inferred rate, together with the corresponding confidence level (hatched regions). Since $\Delta$ and $\rho_{\rm GC}$ are degenerate at fixed $M_{\rm c}$ (cf.~eq.~\eqref{eqn:rate_definition}), we fold them into the single combination
$\rho_{{\rm GC},0}$---the initial mass density of globular clusters.

Stronger constraints arise if one also requires that clusters dominate the rate above $m_1\simeq 50\,M_\odot$.
Under this assumption, we require additionally that the cluster contribution match the observed rate, integrated over this mass range. Indeed, the resulting allowed parameter space (coloured regions in Fig.~\ref{fig:constraints summary 1}) is reduced substantially.

The exclusions are already significant. Models with combinations of $M_{\rm c}$ and $\rho_{\rm h,i}$, that retain merger remnants efficiently and enhance repeated mergers, are constrained most strongly, because they produce too many high-mass events. Likewise, increasing $\rho_{{\rm GC},0}$ raises the overall rate normalisation and enlarges the excluded region. The allowed region is non-trivial, but does include values of $M_{\rm c}$ inferred from globular-cluster observations~\cite{Jordanetal2007,AntoniniGieles2020b}, thus indicating that the GW merger-rate constraints select a physically plausible region of cluster parameter space, rather than requiring implausible initial conditions. The resulting upper limits on $\rho_{{\rm GC},0}$ are likewise consistent with those inferred from cosmological simulations and the galaxy--halo connection~\cite[][and references therein]{Rodriguezetal2015,AntoniniGieles2020b}. Even without assuming that clusters dominate the observed GW population, GW data therefore already provide a way to constrain---and in some regions exclude---models of globular-cluster formation. In the SM we translate these constraints into constraints on $\rho_{\rm GC}$ by fixing $\Delta$.

\paragraph{Discussion.}
Our cluster models show that dynamically assembled binaries can reproduce the observed gravitational-wave population over a broad mass range. For realistic cluster parameters, they match the merger-rate distribution over $m_1\in[20,100]\,M_\odot$, consistent with the spin-based inference from high-mass events, $m_1\gtrsim 50\,M_\odot$~\cite{2022ApJ...941L..39W, 2023arXiv230302973L,Antoninietal2025a,Antoninietal2025b,Antoninietal2025c,HuiMaya,RayKalogera2026,Rayetal2026}.
If 2G black holes from clusters do account for the observed rate above a threshold mass $m_\ast$, these models necessitate the existence of a minimum accompanying 1G contribution below $m_\ast$. Taking $m_\ast=50\,M_\odot$, we minimise $\mathcal{R}(m_1\leq 50\,M_\odot)$ over all models that (i) reproduce $\mathcal{R}(m_1\geq 50\,M_\odot)$ within the 99\% confidence interval inferred from the GW data, and (ii) do not exceed the inferred 99$^{\rm th}$ percentile below $50\,M_\odot$.\footnote{The minimisation is performed over all models considered in this work, both in the main text and in the SM.} This yields
$\mathcal{R}(m_1\leq 50\,M_\odot)\geq 0.099\,{\rm Gpc}^{-3}\,{\rm yr}^{-1},$
from clusters, at 99\% confidence. Thus, dynamical formation in clusters provides a viable explanation for a significant fraction of the observed merger population, and for most mergers with a primary mass of $\simeq 35M_\odot$.

Importantly, however, our main conclusion does not rely on the details of this modelling. The emergence of a second peak near $70\,M_\odot$ is a largely model-independent prediction: if the $\sim 35\,M_\odot$ feature is produced by clusters and a PISN mass gap is present, then cluster mergers inevitably generate a 2G peak at roughly twice the 1G mass scale. Identifying such a feature in the data robustly would therefore have direct implications for our understanding of black-hole formation.

Independently of how much clusters do contribute in reality to the LVK rate, we showed that GW observations allow to constrain globular-cluster populations at birth, in particular the density of clusters in the Universe, $\rho_{\rm GC}$, the Schechter mass $M_{\rm c}$ and the initial half-mass density of globular clusters $\rho_{\rm h,i}$. Thus, GW observations give us a unique window into the conditions for cluster formation, and to events occurring over $10$ Gyr ago, which are poorly understood from electromagnetic observations. 

The black-hole mass distribution predicted by our cluster models exhibits a clear suppression of binaries with component masses below $\sim 20~M_\odot$ (Fig.~\ref{fig:example_rates}). By contrast, Ref.~\cite{Ye_etal_2026} reported a distribution that peaks near $\sim 8~M_\odot$. We believe this discrepancy arises mostly because their analysis 
assumed that the cluster formation-time distribution $\phi(t_{\rm f})$ directly follows the cosmic star‑formation rate.
This might be difficult to reconcile with the tendency of globular clusters to form at high redshift~\citep{Badryetal2019}, and with their markedly different formation and evolution relative to open clusters, which dominate the local star-formation rate~\cite{HeggieHut2003,Binney}.
Ref.~\cite{FishbackFragione2023} relied on the {\tt CMC} catalogue~\cite{Kremeretal2020} and fixed $\rho_{\rm GC}$, whereas our cluster suite is orders of magnitude larger and treats $\rho_{\rm GC}$ as a free parameter. Moreover, Refs.~\cite{FishbackFragione2023,Ye_etal_2026} adopted a metallicity relation tied to the overall galaxy population~\cite{MadauFragos2017}, whereas here we use one calibrated specifically to globular clusters (see SM).
Ref.~\cite{FishbackFragione2023} constrained globular-cluster populations using GWTC-3 data, but under the assumption that the full isotropic-spin population, including 1G mergers, originates in clusters. We do not require this: indeed, our first set of constraints (Fig.~\ref{fig:constraints summary 1}, yellow) is fully independent of whether any observed GW events come from clusters, while the second requires only that 2G mergers above the PISN gap originate in clusters, consistent with their spin population properties~\cite{2022ApJ...941L..39W,Antoninietal2025a,Antoninietal2025b}.

\paragraph{Acknowledgements.}
We thank Thomas Callister for providing an initial version of the code on which our population inference analysis of the gravitational-wave data was based, and thank him, Ignacio Maga\~{n}a Hernandez, Bence Kocsis, and Antonella Palmese for for many insightful discussions. We thank Mario Spera for reviewing an earlier version of this work and for providing useful comments. 
This work was supported by the United Kingdom’s Science and Technology Facilities Council (STFC) grants ST/W000903/1, ST/V005618/1 and UKRI2489, and by a Leverhulme Trust International Professorship Grant (No.~LIP-2020-014). Y.B.G.'s work was partly supported by the Simons Foundation via a Simons Investigator Award to A.A.~Schekochihin (No.~930121). M.G.~acknowledges support from grants PID2024-155720NB-I00 and CEX2024-001451-M by the Spanish Ministry of Science and Innovation and the Spanish State Research Agency (MCIN/AEI/10.13039/501100011033).

This research has made use of data or software obtained from the Gravitational Wave Open Science Center (gwosc.org), a service of the LIGO Scientific Collaboration, the Virgo Collaboration, and KAGRA. This material is based upon work supported by NSF's LIGO Laboratory which is a major facility fully funded by the National Science Foundation, as well as the STFC, the Max-Planck-Society (MPS), and the State of Niedersachsen/Germany for support of the construction of Advanced LIGO and construction and operation of the GEO600 detector. Additional support for Advanced LIGO was provided by the Australian Research Council. Virgo is funded, through the European Gravitational Observatory (EGO), by the French Centre National de Recherche Scientifique (CNRS), the Italian Istituto Nazionale di Fisica Nucleare (INFN) and the Dutch Nikhef, with contributions by institutions from Belgium, Germany, Greece, Hungary, Ireland, Japan, Monaco, Poland, Portugal, Spain. KAGRA is supported by Ministry of Education, Culture, Sports, Science and Technology (MEXT), Japan Society for the Promotion of Science (JSPS) in Japan; National Research Foundation (NRF) and Ministry of Science and ICT (MSIT) in Korea; Academia Sinica (AS) and National Science and Technology Council (NSTC) in Taiwan.

\title{Second-Generation Mass Peak in the Gravitational-Wave Population Probes Globular Clusters}

\bibliography{clusters}

\newpage

\clearpage
\setcounter{equation}{0}
\setcounter{figure}{0}
\setcounter{table}{0}
\renewcommand{\theequation}{S\arabic{equation}}
\renewcommand{\thefigure}{S\arabic{figure}}
\renewcommand{\thetable}{S\arabic{table}}

\setcounter{page}{1}
\appendix
\onecolumngrid
\section{Supplementary Material}\label{SM}
\date{\today}

\subsection{Transfer of first-generation mass features to the hierarchical-merger population}
\label{sec:transfer_kernel}
We argue in the main text that the feature near $m_1\sim 70\,M_\odot$ is not simply the result of doubling the characteristic 1G mass scale. Let us now make this statement more precise by introducing an effective transfer-kernel picture for how a structure in the 1G distribution is mapped onto the population of 2G black holes above the pair-instability scale.

In what follows, $f_{\rm 1G}(m)$ denotes the population-averaged mass distribution of 1G mergers from globular clusters. The corresponding 2G distribution generated by $f_{\rm 1G}$ should likewise be understood as a population-averaged quantity. The pairing, remnant-mass, and retention kernels introduced below are therefore to be interpreted as effective kernels, averaged over the cluster population.

For dynamically assembled binaries, the probability of selecting a mass ratio $q\equiv m_2/m_1\le 1$ at fixed primary mass $m_1$ can be approximated as \cite{Antoninietal2023}
\begin{equation}\label{eq:pairing_kernel}
    p(q\mid m_1)\propto q^{\alpha_2},
\end{equation}
with $\alpha_2=3.5+\alpha$, where $\alpha$ is the slope of the underlying black-hole mass function. This form captures the tendency of cluster dynamics to favour relatively large mass ratios \cite{AarsethHeggie1976,Atallahetal2024,GinatPerets2024}. More generally, if the 1G feature is narrow, then even relatively generic pairing prescriptions yield mergers with characteristic $q\sim 1$, so the detailed pairing kernel mainly broadens the mapped 2G feature rather than strongly shifting its centroid. Sensitivity to the pairing distribution becomes important only if the 1G peak is broad.

The population-averaged mass distribution of 2G merger remnants may then be written as
\begin{equation}
    g_{\rm 2G}(M)
    =
    \int \mathrm{d}m_1 \int \mathrm{d}q\;
    f_{\rm 1G}(m_1)\,
    p(q\mid m_1)\,
    P_{\rm ret}(m_1,q,\chi)\,
    P_{\rm f}(M\mid m_1,q,\chi),
    \label{eq:g2_general_dyn}
\end{equation}
where $P_{\rm ret}(m_1,q,\chi)$ is the probability that the merger remnant is retained by the cluster, $P_{\rm f}(M\mid m_1,q,\chi)$ is the remnant-mass kernel, and $\chi$ denotes the black-hole spin parameters entering the remnant-mass and recoil prescriptions.

Approximating the remnant mass as
\begin{equation}
    M \simeq \eta(q,\chi)\,(1+q)m_1,
    \label{eq:remnant_mass}
\end{equation}
where $\eta(q,\chi)$ denotes the fraction of the total binary mass retained after GW emission, Eq.~\eqref{eq:g2_general_dyn} becomes
\begin{equation}
    g_{\rm 2G}(M)
    \propto
    \int \mathrm{d}m_1 \int \mathrm{d}q\;
    f_{\rm 1G}(m_1)\,
    q^{\alpha_2}\,
    P_{\rm ret}(m_1,q,\chi)\,
    \delta\!\left[M-\eta(q,\chi)(1+q)m_1\right].
    \label{eq:g2_delta_dyn2}
\end{equation}
Performing the $m_1$ integral yields
\begin{equation}
    g_{\rm 2G}(M)
    \propto
    \int \mathrm{d}q\;
    f_{\rm 1G}\!\left(\frac{M}{\eta(q,\chi)(1+q)}\right)\,
    q^{\alpha_2}\,
    \frac{
        P_{\rm ret}\!\left(\frac{M}{\eta(q,\chi)(1+q)},q,\chi\right)
    }{
        \eta(q,\chi)(1+q)
    }.
    \label{eq:g2_transfer}
\end{equation}
Equation~\eqref{eq:g2_transfer} shows explicitly that the 2G population is obtained by mapping the 1G merger distribution upward in mass through an effective transfer kernel set by pairing, remnant-mass losses, and retention.

To illustrate this mapping let us consider a localised feature in the population-averaged 1G distribution centred at $m_0$ with width $\sigma_m$.
We assume that the effective pairing distribution is narrow, so that $q$ is concentrated near some characteristic value $\bar q$, and that both $\eta_{\rm  }$ and $P_{\rm ret}^{\rm  }$ vary slowly across the support of the 1G feature. 
Then,
the corresponding 2G feature is centred at 
\begin{equation}
    M_0 \simeq \eta_{\rm  }(\bar q)\,(1+\bar q)\,m_0 
    \label{eq:M0_transfer_dyn}
\end{equation}
For near-equal-mass mergers and $\eta(1)\simeq 0.95$, one finds $M_0\simeq 1.9\,m_0$, so a feature near $35\,M_\odot$ is mapped to $\sim 67\,M_\odot$. The width of the transferred feature is broadened by both the intrinsic width of the 1G feature and the spread in effective mass ratios:
\begin{equation}
    \sigma_M^2 \simeq
    \left[\eta_{\rm  }(\bar q)(1+\bar q)\right]^2 \sigma_m^2
    +
    m_0^2
    \left[
    \eta_{\rm  }(\bar q)+(1+\bar q)\eta_{\rm  }'(\bar q)
    \right]^2
    \sigma_q^2
    +
    \sigma_{\rm rem}^2,
    \label{eq:sigmaM_transfer_dyn}
\end{equation}
where $\sigma_q$ is the width of the effective pairing distribution and $\sigma_{\rm rem}$ accounts for additional scatter from spins and remnant-mass fitting uncertainties.

For instance, if we approximate the first-generation merger feature by a Gaussian,
\begin{equation}
    f_{\rm 1G}(m)=\frac{1}{\sqrt{2\pi}\sigma_{\rm 1G}}
    \exp\!\left[-\frac{(m-\mu_{\rm 1G})^2}{2\sigma_{\rm 1G}^2}\right],
    \label{eq:f1g_gaussian}
\end{equation}
where $\mu_{\rm 1G}$ and $\sigma_{\rm 1G}$ are the mean and width of the 
1G feature, then the 2G remnant distribution is 
\begin{equation}
    g_{\rm 2G}(M)
    \propto
    \frac{1}{\eta_{\rm  }(\bar q)(1+\bar q)}
    \exp\left\{
    -\frac{\left[\frac{M}{\eta_{\rm  }(\bar q)(1+\bar q)}-\mu_{\rm 1G}\right]^2}{2\sigma_{\rm 1G}^2}
    \right\} ,
    \label{eq:g2_gaussian_narrowq}
\end{equation}
with mean
$
\mu_{\rm 2G}\simeq \eta_{\rm  }(\bar q)(1+\bar q) \mu_{\rm 1G}\,, 
\label{eq:mu2g}
$
and variance 
$
\sigma_{\rm 2G}^2\simeq \eta(\bar q)^2(1+\bar q)^2\sigma_{\rm 1G} ^2 \,.
$

This mapping has two important implications. First, the centroid and width of the transferred feature are set jointly by the characteristic 1G mass scale, the remnant-mass fraction $\eta_{\rm  }$, and the dynamical pairing process. Because cluster dynamics favours near-equal-mass 1G mergers, remnant masses are deposited near a relatively narrow range around $\eta_{\rm }(1+\bar q)m_0$, rather than being dispersed into a broad high-mass tail. The location of the second feature is therefore relatively robust, while its width encodes the pairing statistics and remnant-mass losses. Second, its normalisation depends on the efficiency of remnant retention and recycling, and is therefore sensitive to the cluster population.

We note that for a narrowly peaked 1G feature, the detailed pairing prescription is not especially important: it mainly broadens the 2G feature and causes at most a modest shift in its centroid. If the 1G population is narrowly concentrated around the first peak, then most mergers contributing to the remnant population will naturally have $q\sim 1$, even without strong dynamical preference for equal-mass pairing. Sensitivity to the pairing distribution only becomes significant if the 1G peak is broad, for in that case mergers can draw components from a wider range of masses.

The observed feature in the primary-mass distribution is expected to be dominated by $2{\rm G}+1{\rm G}$ mergers. In this regime the primary mass is typically the remnant mass itself, so that
\begin{equation}
    \frac{\mathrm{d}\mathcal{R}_{2{\rm G}+1{\rm G}}}{\mathrm{d}m_1}
    \propto g_{\rm 2G}(m_1)~.
    \label{eq:m1_g2g_dyn}
\end{equation}
The feature near $70\,M_\odot$ in the primary-mass spectrum therefore directly traces the transferred remnant-mass feature. In writing Eq.~\eqref{eq:m1_g2g_dyn}, we have assumed that the probability for a retained 2G black hole to merge again (relative to other 2G black holes) is approximately independent of its mass. This is a good approximation in the cluster models considered here, since each cluster typically hosts at most one 2G black hole at a time, and that object is the most massive black hole in the cluster. Thus as a result of equation~\eqref{eq:pairing_kernel} any 2G black hole is expected to enter a subsequent merger rapidly, so that later dynamical selection should not distort the shape of $g_{\rm 2G}(m_1)$ significantly.

Finally, the visibility of this structure depends not only on the transfer kernel but also on the background against which it is observed. Above the PISN scale the 1G population is strongly depleted, so the total high-mass merger rate may be written schematically as
\begin{equation}
    \mathcal{R}_{\rm high}(M)\approx
    \mathcal{R}_{\rm 1G}(M)+\mathcal{R}_{\rm 2G}(M),
\end{equation}
with $\mathcal{R}_{\rm 1G}(M)$ small. As a result, even a moderately broadened transferred distribution can appear as a distinct second peak rather than as a smooth high-mass tail.

\subsection{Details of cluster probability distributions}
\label{appendix:cluster populations}
The rate in equation \eqref{eqn:rate_definition}, involves the probability distributions of the initial cluster metallicity, formation time, and mass, collectively denoted by $\phi$. We follow ref.~\cite{Antoninietal2023} in determining them. Explicitly, $\phi_t(t_{\rm f})$ is based on the distribution of cluster ages in ref.~\cite{Badryetal2019}. For $\phi_Z(Z|t_{\rm f})$ we use a log-normal distribution, inferred from cluster data \cite{VandenBerg2013}, given by 
\begin{equation}
    \phi_Z(Z|t_{\rm f}) = \frac{\log_{10}(e)}{\sqrt{2\pi \sigma^2}} \exp\left[ -\frac{\left(\log_{10}(Z/Z_{\odot}) - \log_{10}\left(\overline{Z}(t_{\rm f})/Z_{\odot}\right)\right)^2}{2\sigma^2}\right]\,,
\end{equation}
where the Solar metallicity is $Z_{\odot} = 0.0134$ \cite{Asplundetal2009}, $\sigma = 0.25$, and 
\begin{equation}
    \log_{10}\left(\frac{\overline{Z}(t_{\rm f})}{Z_{\odot}}\right) = 0.42267 + 0.04559\left(\frac{t_{\rm f}}{\textrm{Gyr}}\right) - 0.017 \left(\frac{t_{\rm f}}{\textrm{Gyr}}\right)^2\,.
\end{equation}
The metallicity also determines the initial mass function of the clusters, with the prescription of ref.~\cite{Badryetal2019} (cf.~\cite{Antoninietal2023}). Initial black hole spins are either set to zero or, in some models (see below), drawn from a truncated Gaussian, \emph{viz.}
\begin{equation}\label{eqn:LVK spin distribution}
    \phi(\chi) = \frac{\mathrm{e}^{-(\chi-\mu_{\chi})^2/2\sigma_{\chi}^2}}{\int_0^1\mathrm{e}^{-(x-\mu_{\chi})^2/2\sigma_{\chi}^2}\mathrm{d}x}\,,
\end{equation}
where $\mu_{\rm chi} = 0.1$ and $\sigma_{\chi} = 0.3$---the mean values of the posterior found by LVK \cite[][fig.~17]{LVK2025GWTC4}, as inferred from the publicly available data \cite{LVK_GWTC4_2025_zenodo}.

While $\phi_Z$ and $\phi_t$ are fixed across our cluster models, the initial cluster mass-function,
\begin{equation}\label{eqn:Schechter mass function}
    \phi_{M,0}(M_{\rm cl,0}) = \frac{2N}{M_{\rm cl,0}^2}\mathrm{e}^{-M_{\rm cl,0}/2M_{\rm c}}
\end{equation}
does depend on the parameter $M_{\rm c}$, the Schechter mass ($N$ is a normalisation). The \emph{current} mass-function is
\begin{equation}\label{eqn:mass function with mass loss}
    \phi_{M}(M_{\rm cl}) = \frac{N}{(M_{\rm cl}+\Delta)^2}\mathrm{e}^{-(M_{\rm cl}+\Delta)/M_{\rm c}}\,,
\end{equation}
which encapsulates the mass-loss parameter $\Delta$. Given equation \eqref{eqn:rate_definition}, it is only $M_{\rm c}$ which affects the relative distribution of primary mass $m_1$, because it sets the relative fraction of high-mass clusters \emph{vs}.~low-mass ones; $\Delta$ only influences the overall normalisation of the rate.  

The number of possible parametrisations of the initial conditions of clusters is infinite. The functional forms we chose here are motivated by observations and theoretical arguments, and correspond to a physically motivated sub-set of such parametrisations. 
For example, we could have considered $\phi_{M,0} \propto \exp(-M/M_{\rm c})/M^{\alpha}$ for some $\alpha$. The observed present-day mass function is consistent with $\alpha = 2$, equation \eqref{eqn:mass function with mass loss} \cite{Jordanetal2007}; furthermore, we fixed $\alpha = 2$ both because there are already many parameters in our models, and because this exponent at low masses should arise here for theoretical reasons similar to those that give rise to the Press--Schechter mass function in cosmology \cite{PressSchechter1974}, which has a similar dependence. 

\subsection{Additional simulation results and analysis}
In this section we present three extensions of the main analysis. First of all, we describe how the models with non-zero initial spins differ from those presented in the main text. Secondly, we translate the constraints derived in the main text from the combination $K\rho_{\rm GC}$ to the present-day globular-cluster mass density $\rho_{\rm GC}$ itself, by fixing $\Delta$ to a conservative value motivated by Milky Way observations. Thirdly, we consider a broader class of cluster models in which the initial half-mass density depends on cluster mass, and assess how this modifies the predicted merger-rate distributions and the resulting constraints on cluster population parameters. 

\paragraph{Spins.} Figs.~\ref{fig:example_rates}--\ref{fig:constraints summary 1} presented rates for initial black-hole spins were zero. However, LVK coalescences are consistent with spins drawn from Eq.~\eqref{eqn:LVK spin distribution}. Even with zero spins, 2G black holes can still acquire large spins, because some of their progenitors' orbital angular momentum is retained by the remnant black holes, and indeed, heavier black holes are correlated with having larger spins \cite{LVK2025GWTC4}. Spins should not influence the 1G merger rate, because their main effect is to determine the GW kick exerted on the 2G black hole as it forms, and hence whether that black holes is retained in the host cluster or not. 

\begin{figure*}
    \centering
    \includegraphics[width=0.49\textwidth]{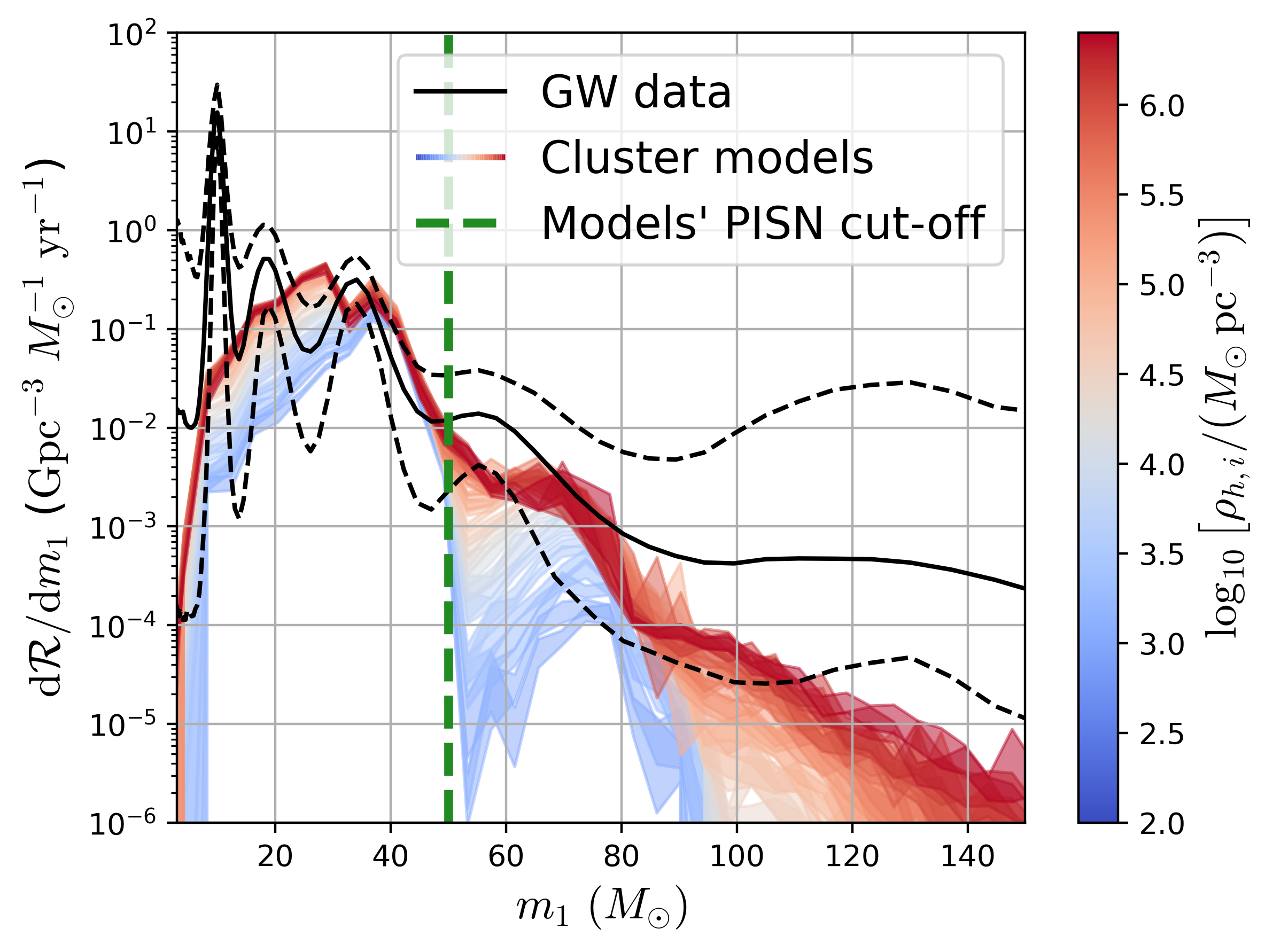}
    \includegraphics[width=0.49\textwidth]{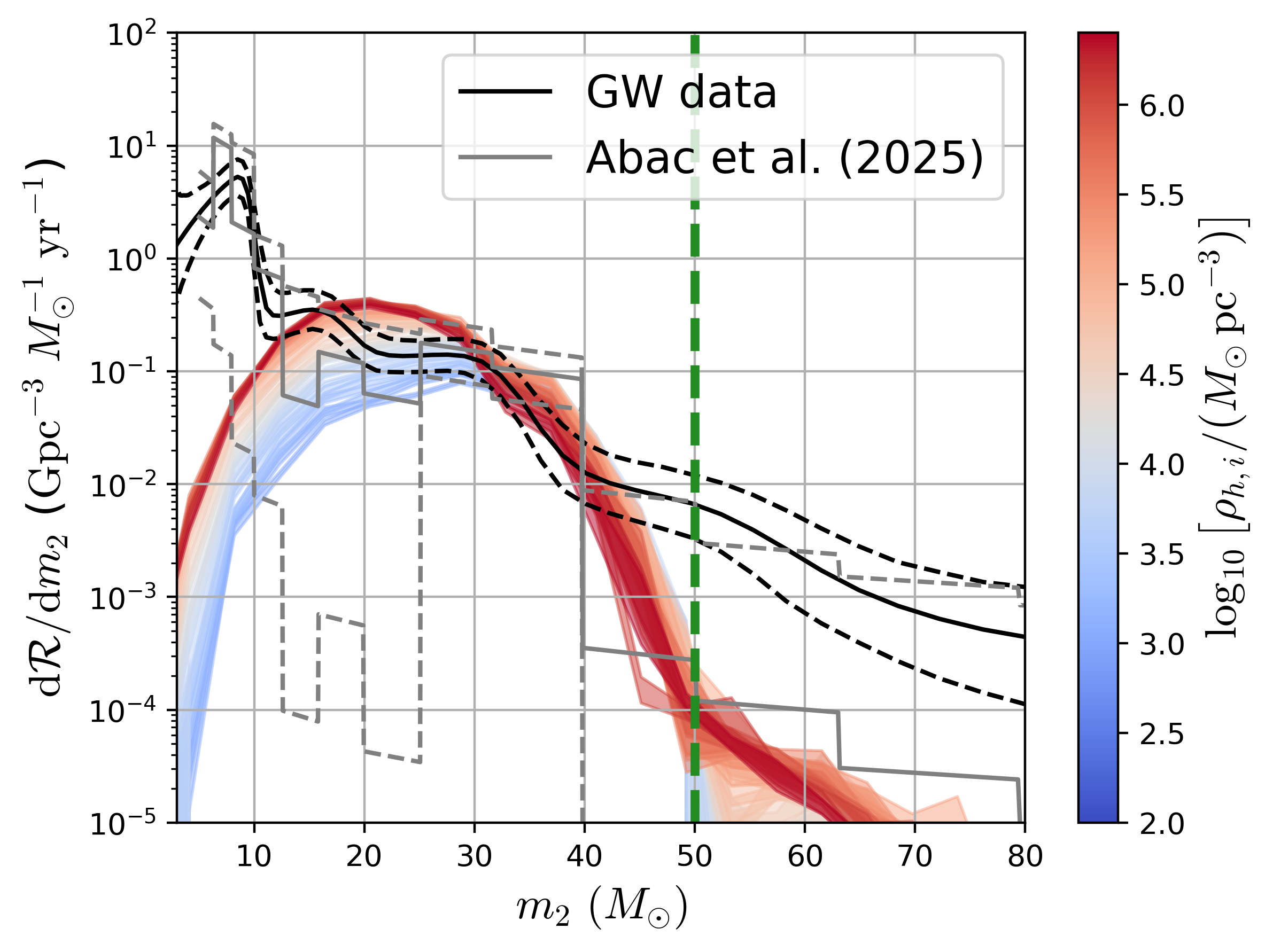}
    \caption{Same as fig.~\ref{fig:example_rates}, but for models where the black holes LVK spins (Eq.~\ref{eqn:LVK spin distribution}), shown for $\rho_{\rm h,i} \geq 10^{3}\,M_{\odot}~\textrm{pc}^{-3}$. \emph{Left}: $\mathrm{d}\mathcal{R}/\mathrm{d}m_1$; \emph{right}: corresponding secondary-mass distribution. The parameters used are $\Delta=10^{5.15}\,M_\odot$, $M_{\rm c}=10^{6.28}\,M_\odot$, and $\rho_{\rm GC}=7\times 10^{14}\,M_\odot\,\mathrm{Gpc}^{-3}$
    }
    \label{fig:example_rates no spins}
\end{figure*}

\begin{figure*}
    \centering
    \includegraphics[width=0.5\textwidth]{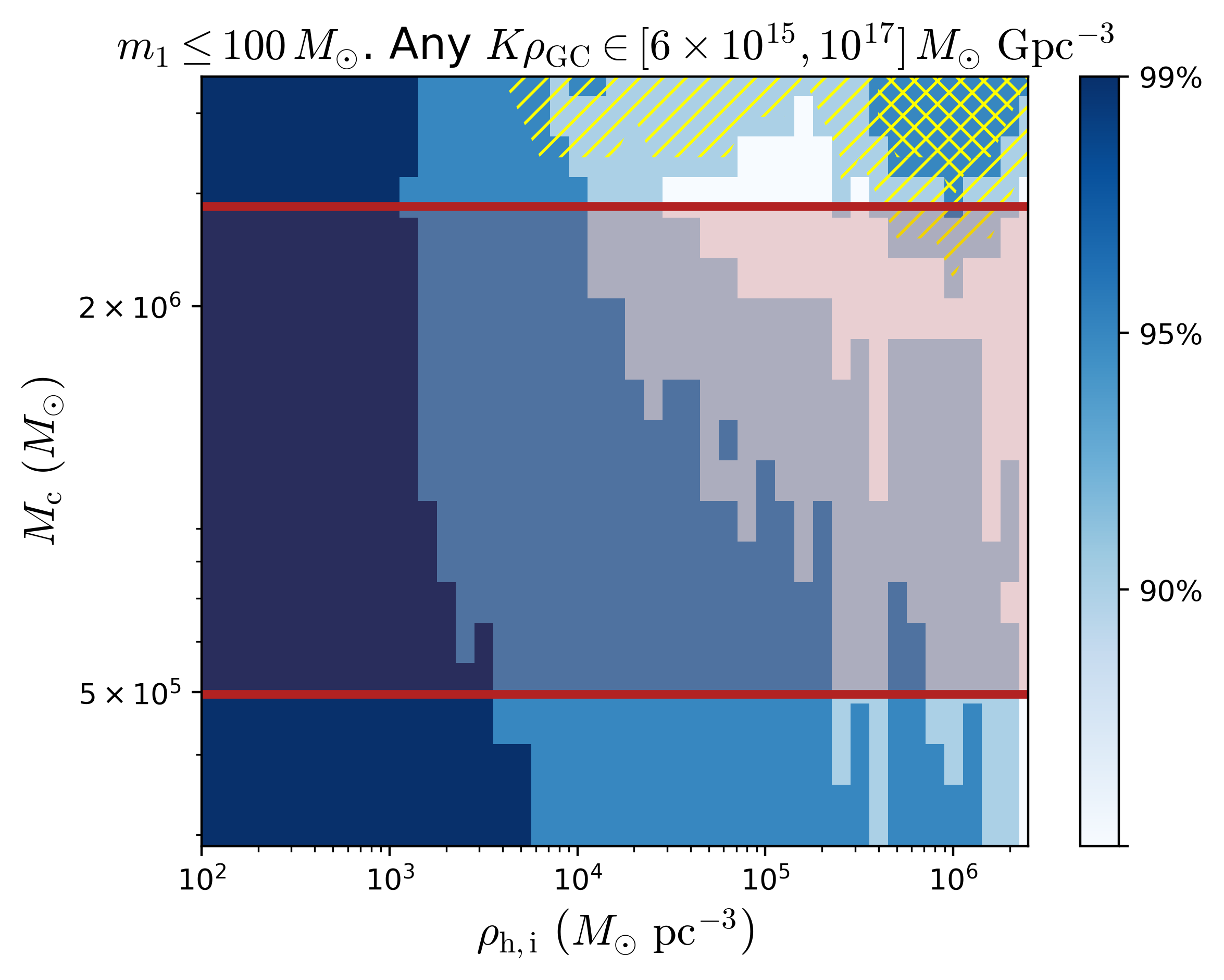}
    \includegraphics[width=0.48\textwidth]{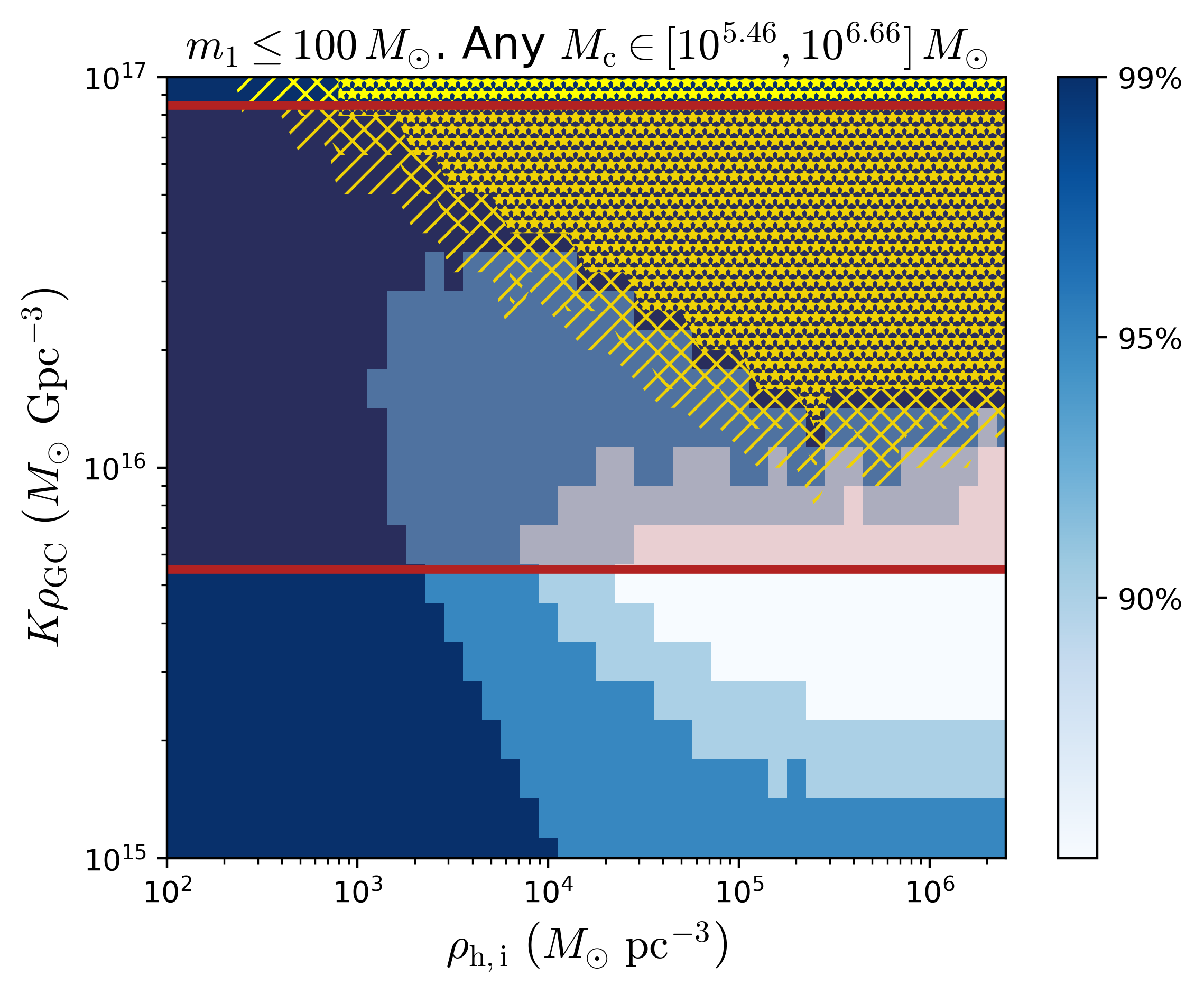}
    \caption{Same as fig.~\ref{fig:constraints summary 1}, but for models where the black holes initially have LVK-distributed spins.}
    \label{fig:constraints summary no spins}
\end{figure*}

Figs.~\ref{fig:example_rates no spins}--\ref{fig:constraints summary no spins} exemplify what changes. We see that the 2G peak is generally higher for models with zero initial spins, for there the GW kick is weaker, and it is easier to retain the merger products, and for them to coalesce again with another 1G black hole. This implies that generally less massive clusters can account for more of the 2G rate, hence modifying constraints on $M_{\rm c}$ and $\rho_{\rm GC}$. 

\begin{figure*}
    \centering
    \includegraphics[width=0.5\textwidth]{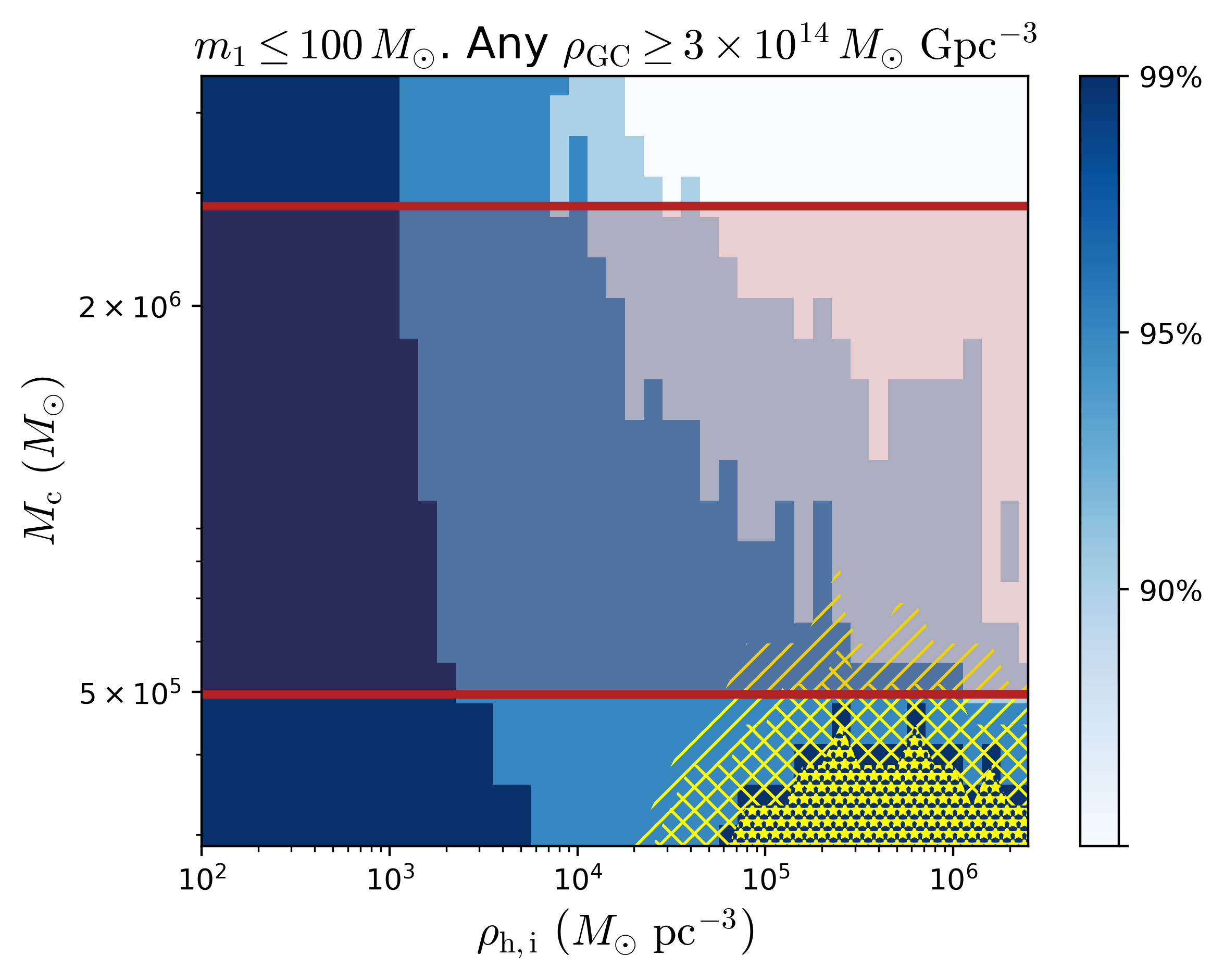}
    \includegraphics[width=0.48\textwidth]{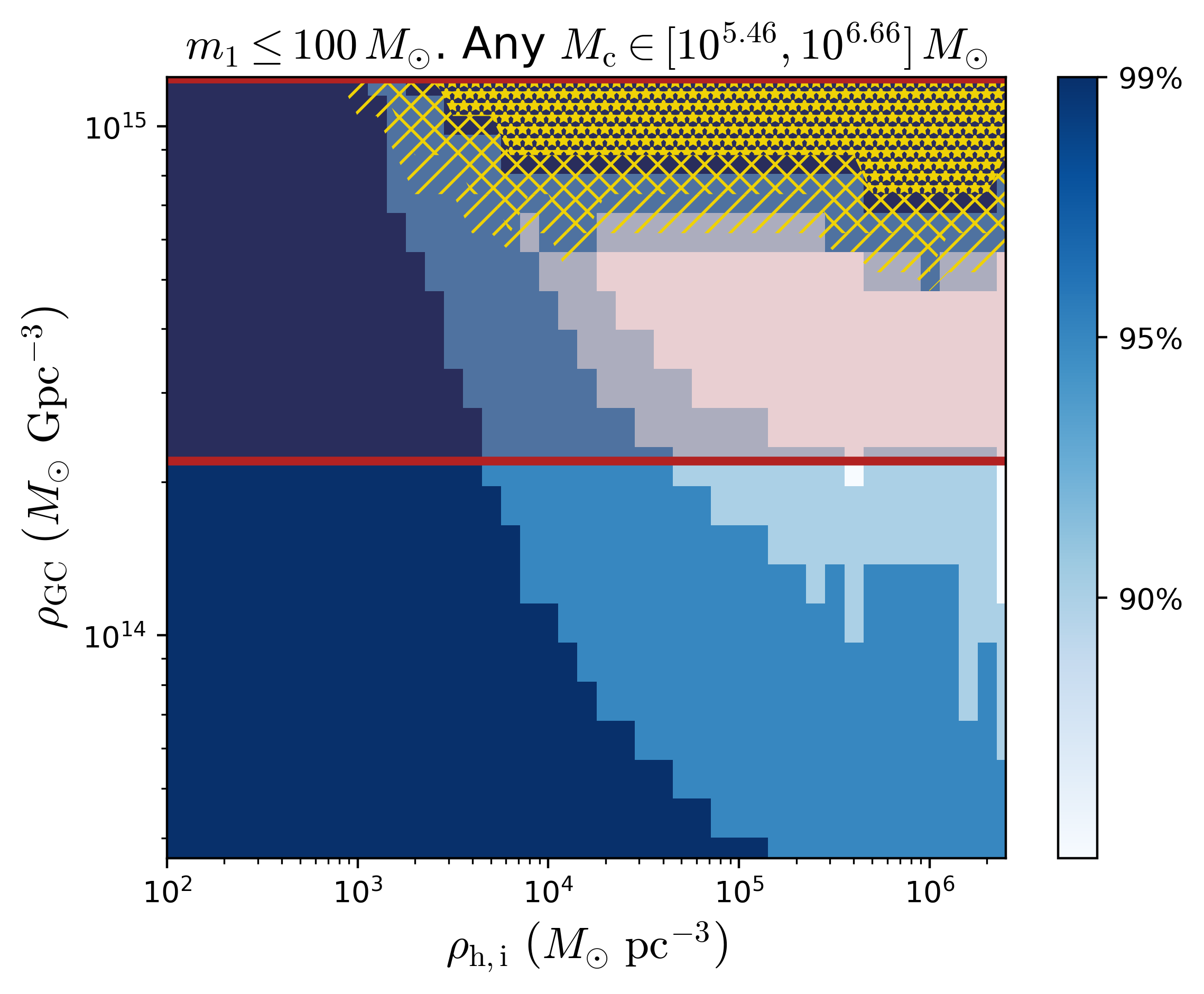}
    \includegraphics[width=0.5\textwidth]{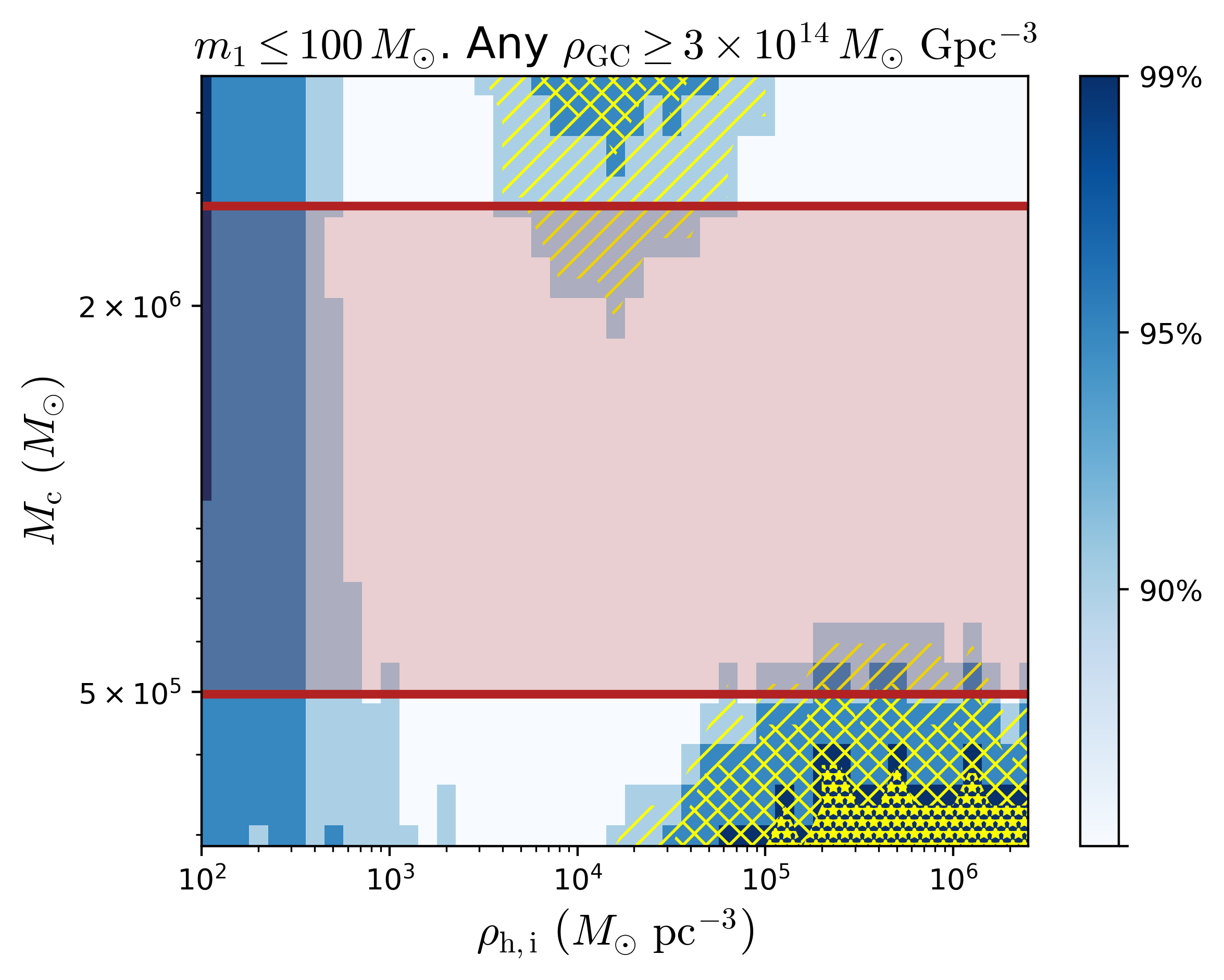}
    \includegraphics[width=0.48\textwidth]{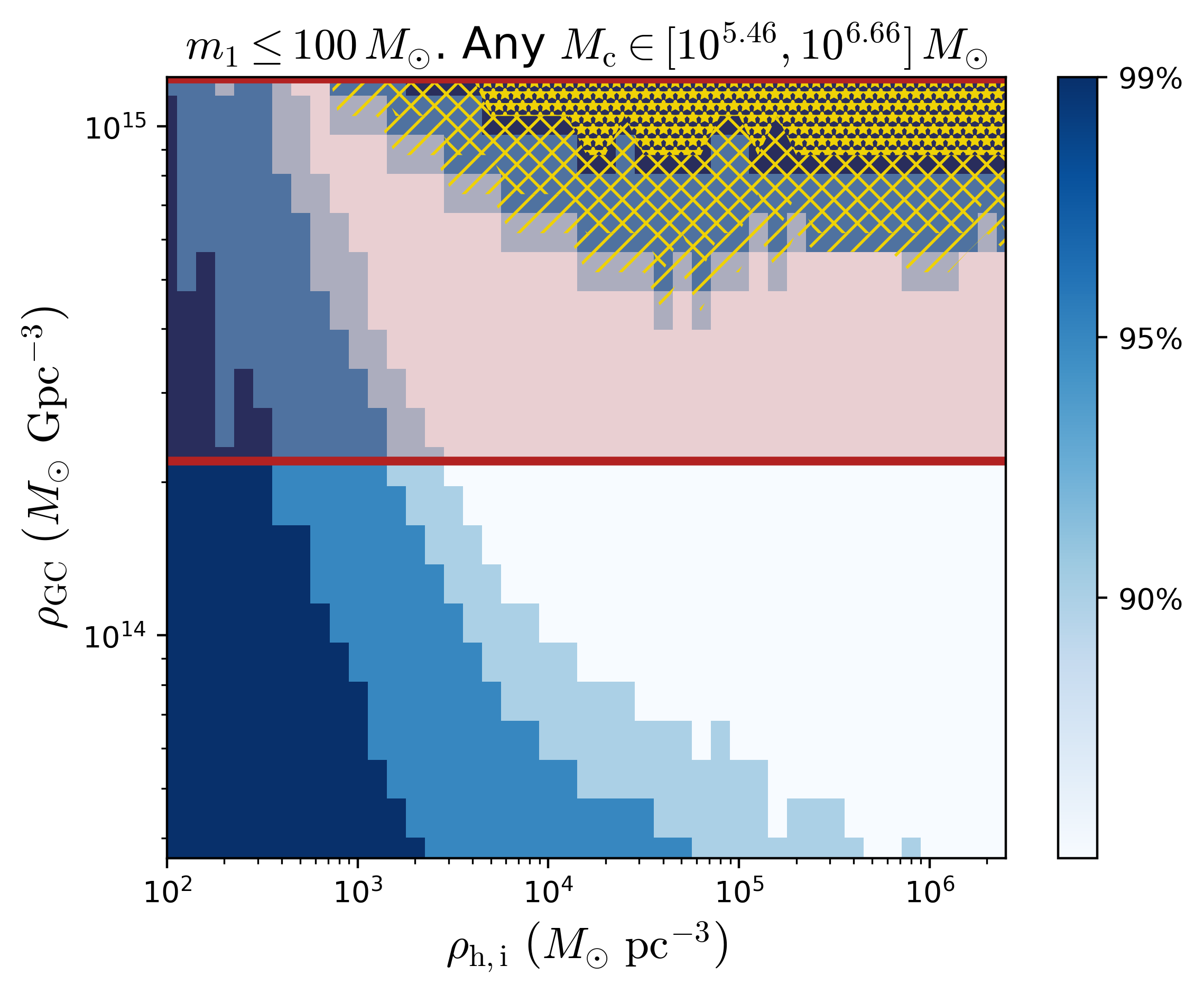}
    \caption{Same as Fig.~\ref{fig:constraints summary 1} in the main text, but for the present-day density of clusters, $\rho_{\rm GC}$, instead of the initial one, $\rho_{\rm GC,0} \equiv K\rho_{\rm GC}$. \emph{Top}: LVK spin distribution (Eq.~\ref{eqn:LVK spin distribution}); \emph{bottom}: zero initial spins.}
    \label{fig:constraints summary 2}
\end{figure*}

\paragraph{Current density of clusters.} Fig.~\ref{fig:constraints summary 1} showed the constraints inferred on the initial mass density of clusters, $\rho_{\rm GC,0}$. Here we present constrains on the present-day density, by assuming a fixed value for $\Delta = 10^{5.15}\,M_{\odot}$, which is the lowest value allowed at $90\%$ from Milky Way observations \cite{Antoninietal2023}. This value minimises $K$ at a fixed $M_{\rm c}$, implying an overall lower rate; thus, the constraints on $\rho_{\rm GC}$ obtained from this value of $\Delta$ are conservative. The result of doing so is shown in Fig.~\ref{fig:constraints summary 2}, which is broadly similar to Figs.~\ref{fig:constraints summary 1} and \ref{fig:constraints summary no spins}, as expected. 

\paragraph{Initial half-mass density.} In the main text we considered a model where all clusters share the same initial half-mass density. While theoretically simple, it is of course unknown whether this held for the initial conditions.\footnote{The local cluster sample in ref.~\cite{2005ApJ...627..203H, 2010MNRAS.408L..16G} is consistent with $\beta\simeq -0.845$ for massive clusters, with substantial scatter.} We thus also consider the case where 
\begin{equation}
    \rho_{\rm h,i} = 10^5 M_{\odot}~\textrm{pc}^{-3}~\left(\frac{M}{10^4 \, M_\odot}\right)^\beta\,,
\end{equation}
for $\beta \in [-1,1]$. Here, the cluster population parameters become $\set{\rho_{\rm GC},M_{\rm c},\beta,\Delta}$. We only consider models with zero initial spins, for simplicity. 

\begin{figure*}
    \centering
    \includegraphics[width=0.49\textwidth]{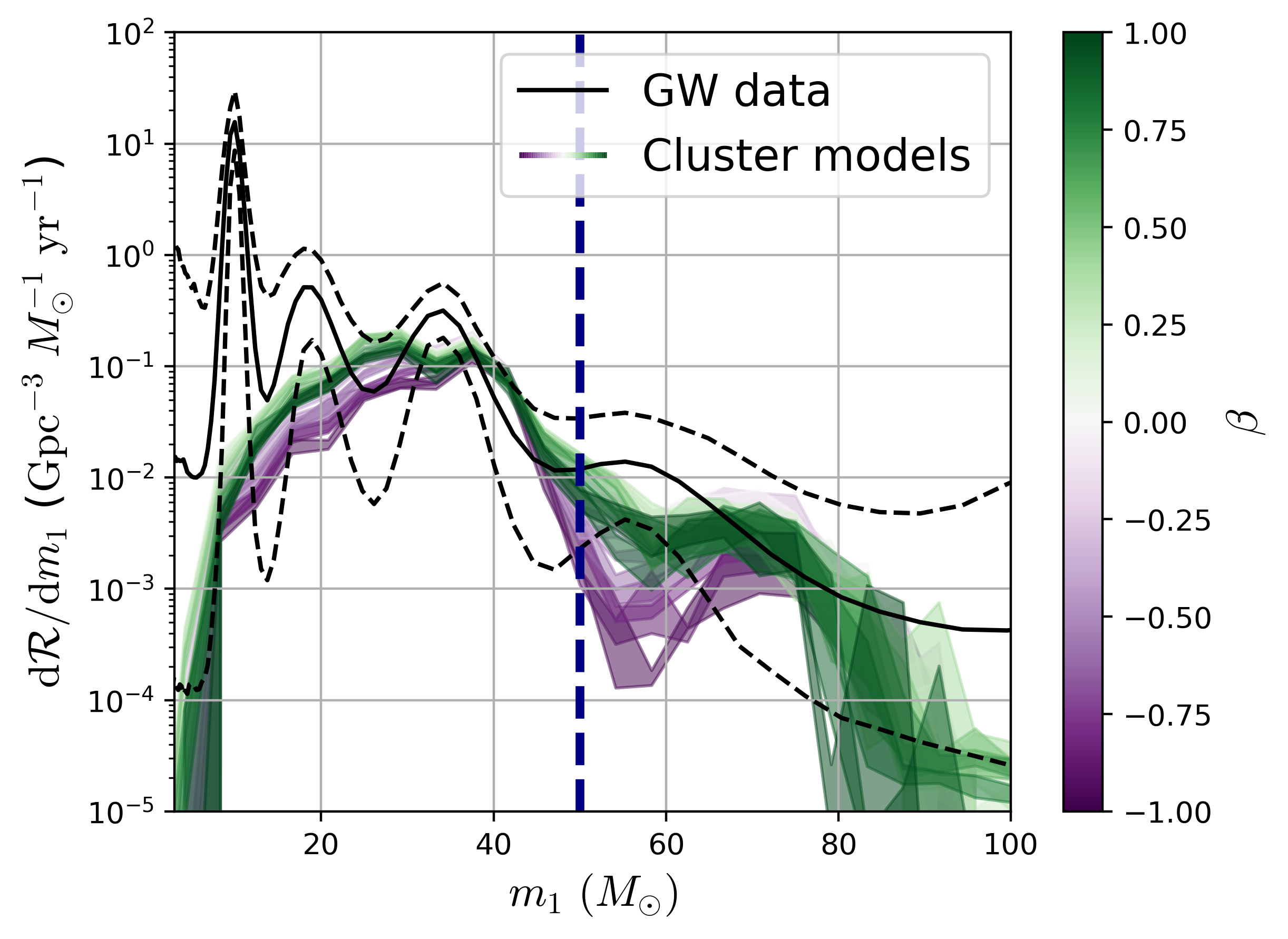}
    \includegraphics[width=0.49\textwidth]{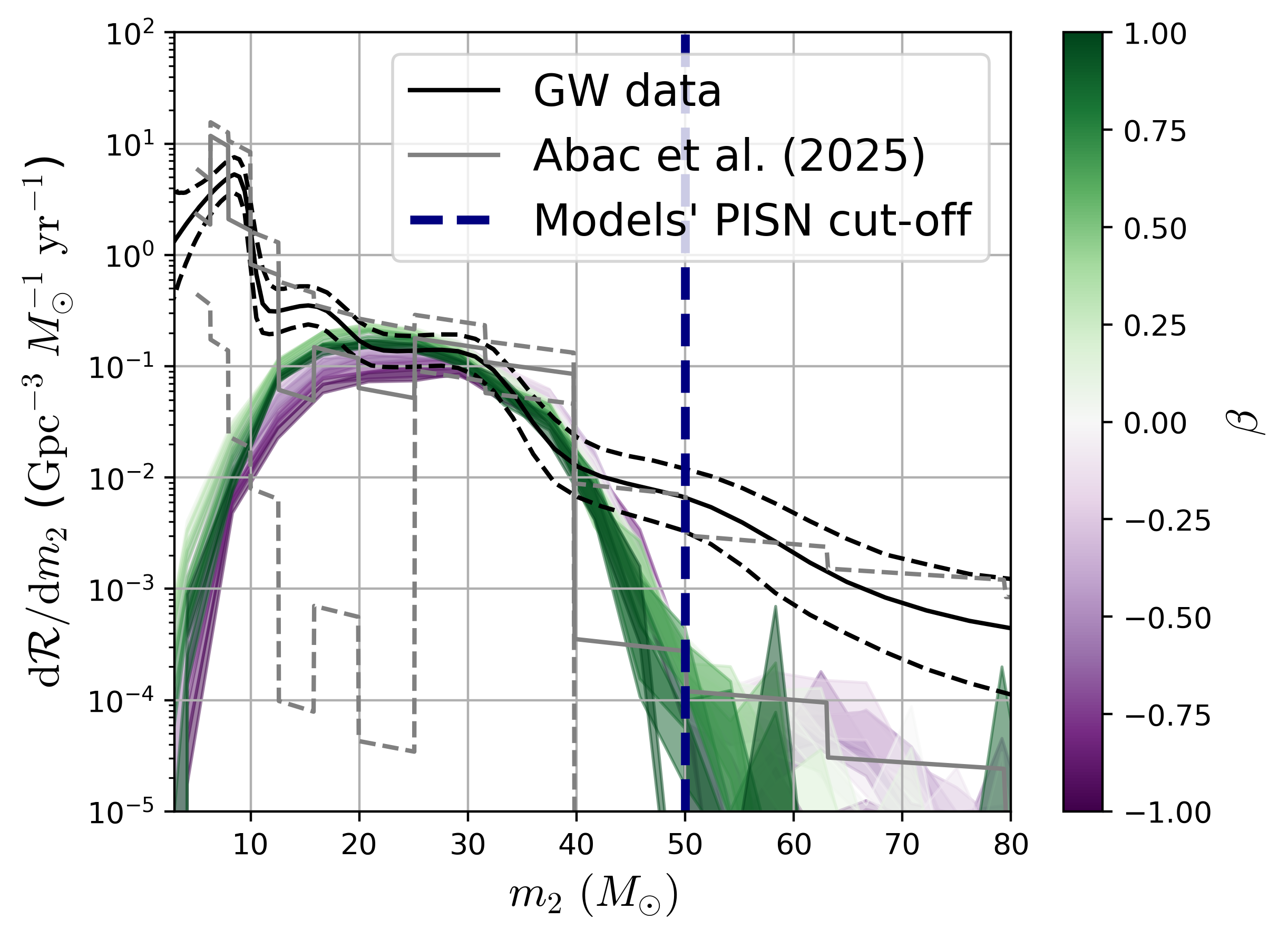}
    \caption{Same as Fig.~\ref{fig:example_rates no spins}, but for models with $\rho_{\rm h,i} \propto M_{\rm cl}^\beta$, as function of the exponent $\beta$ (see text for details).}
    \label{fig:betas}
\end{figure*}

\begin{figure*}
    \centering
    \includegraphics[width=0.49\textwidth]{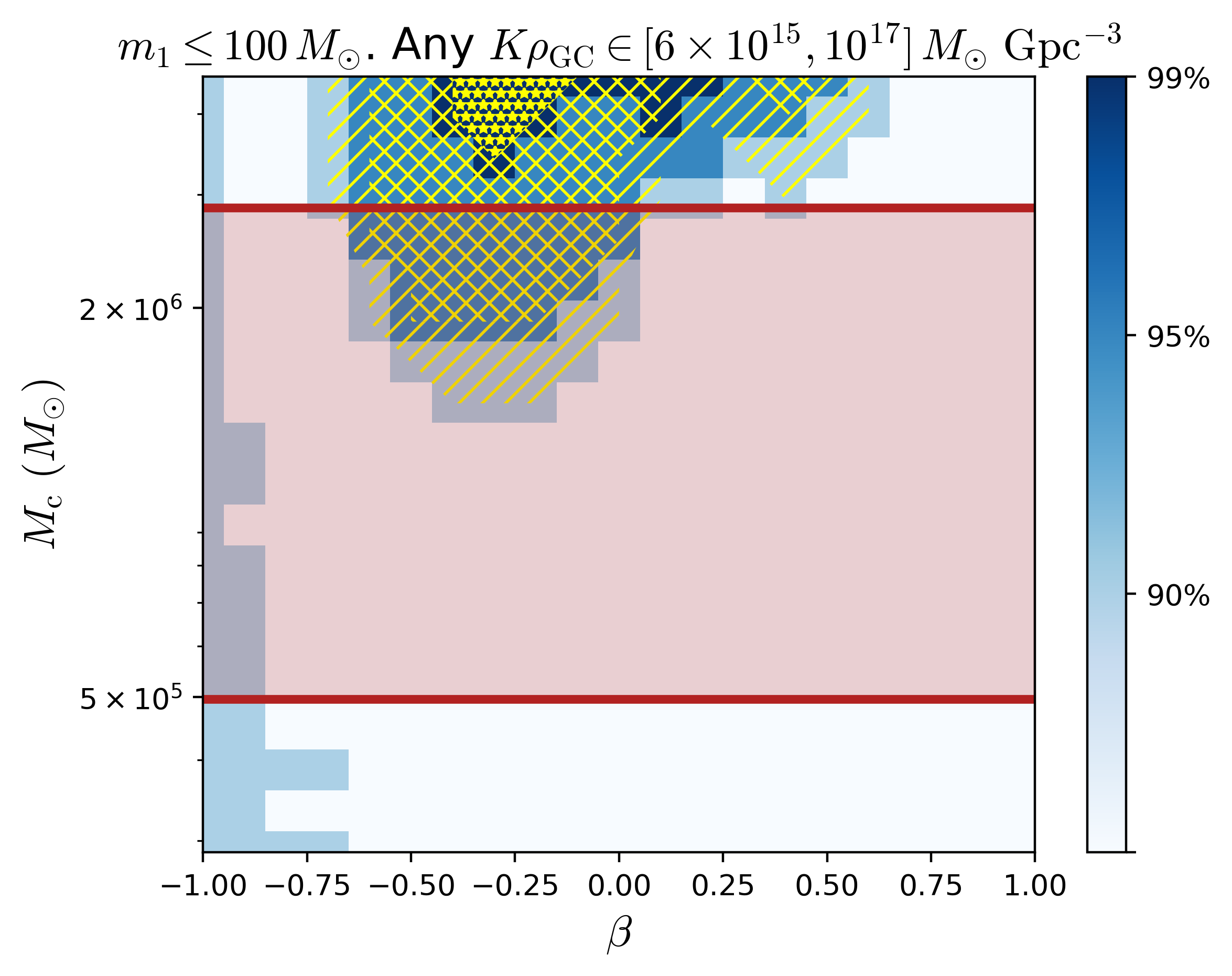}
    \includegraphics[width=0.48\textwidth]{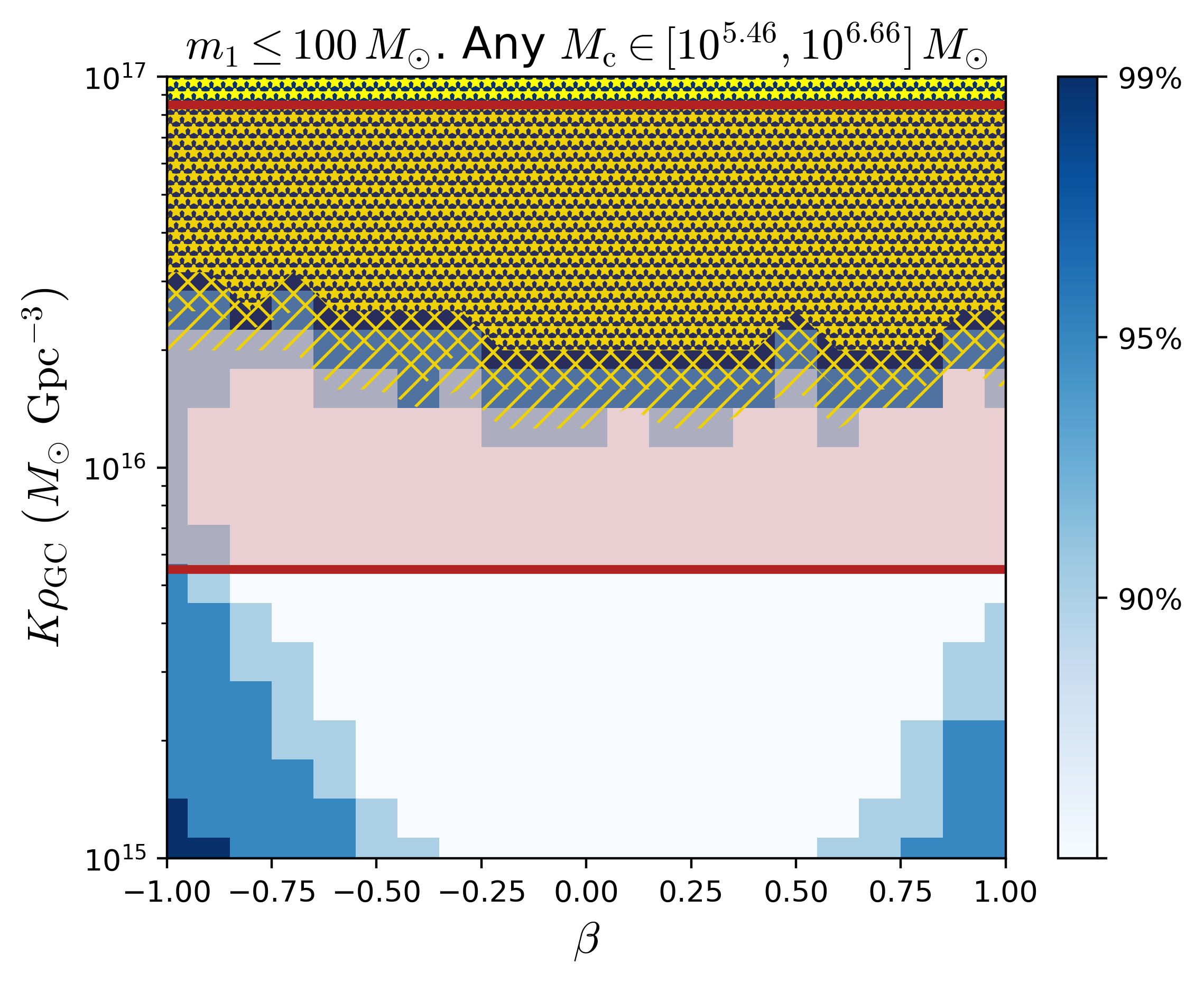}

    \caption{Same as the two panels of Fig. \ref{fig:constraints summary no spins}, but for the $\rho_{\rm h,i} \propto M_{\rm cl}^\beta$ models.}
    \label{fig:constraints betas}
\end{figure*}

\begin{figure*}
    \centering
    \includegraphics[width=0.49\textwidth]{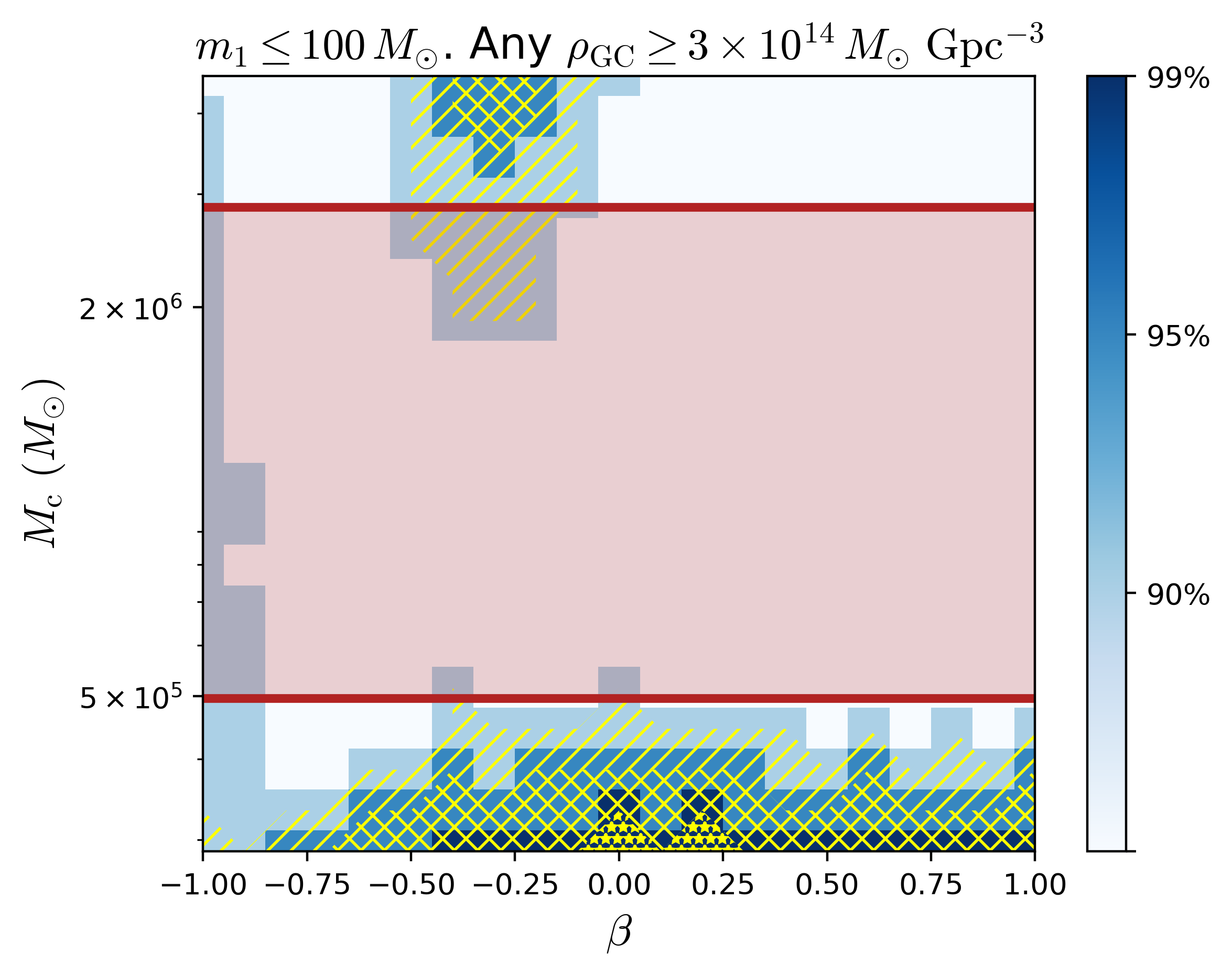}
    \includegraphics[width=0.48\textwidth]{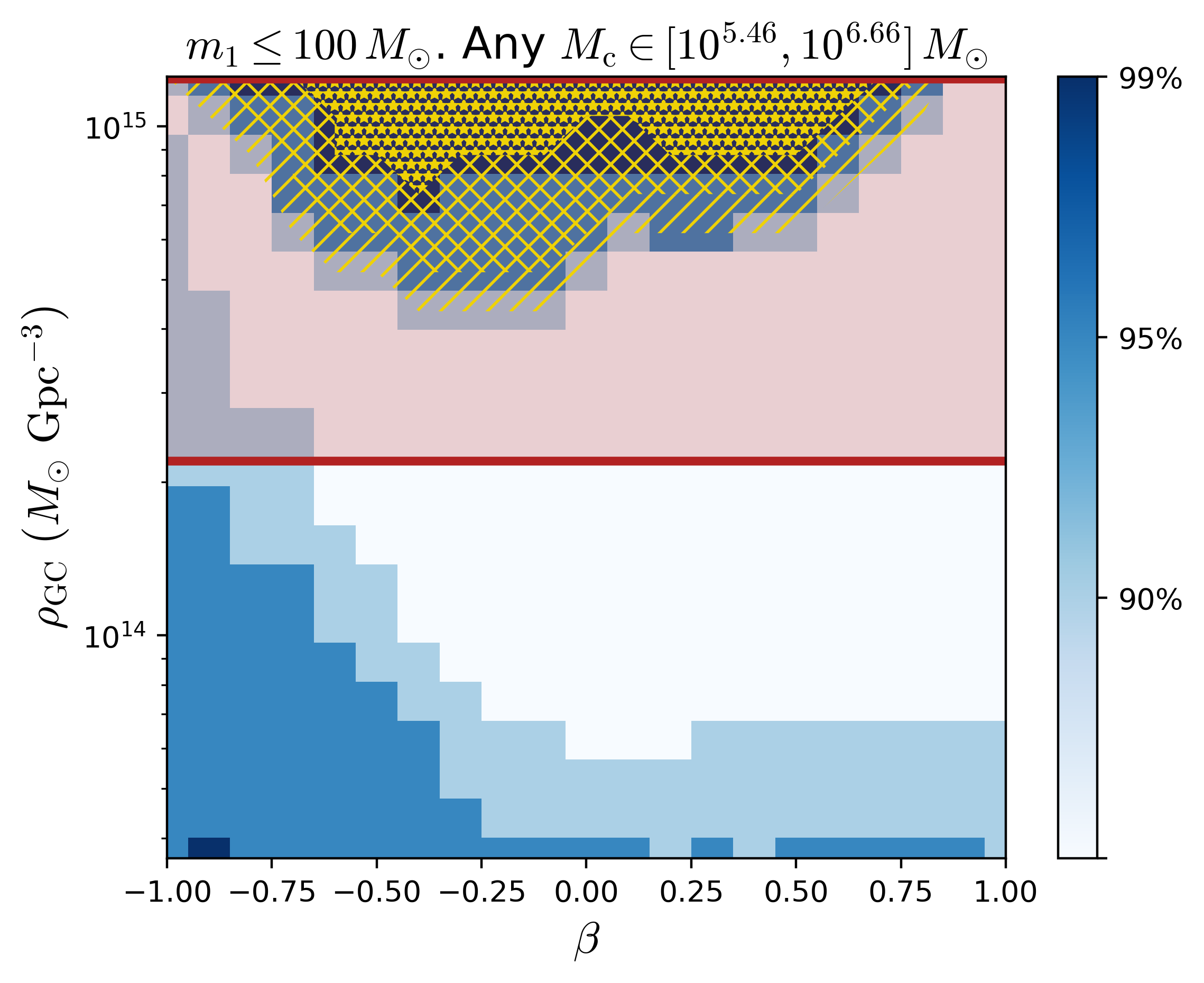}

    \caption{Same as bottom row of Fig. \ref{fig:constraints summary 2}, but for the $\rho_{\rm h,i} \propto M_{\rm cl}^\beta$ models.}
    \label{fig:constraints betas 2}
\end{figure*}

In Fig.~\ref{fig:betas} we show what such models predict for $\mathrm{d}\mathcal{R}/\mathrm{d}m_1$ and $\mathrm{d}\mathcal{R}/\mathrm{d}m_2$. The 2G peak in the $m_1$ distribution is present here, too; it is more prominent for $\beta < 0$. Similarly to the uniform-$\rho_{\rm h,i}$ case, the $m_2$ distribution does not have a strong 2G peak, and does decay rapidly, but smoothly, at $\gtrsim 40\,M_{\odot}$. 
The equivalents of Figs.~\ref{fig:constraints summary 1} and \ref{fig:constraints summary 2} for these models is shown in Figs.~\ref{fig:constraints betas} and \ref{fig:constraints betas 2}, respectively. Qualitatively, they are similar to those from the constant-$\rho_{\rm h,i}$ model, in-as-much as they are able to place an upper limit on $\rho_{\rm GC}$ within these models. Positive values of $\beta$ are slightly preferred to negative ones.

\subsection{Population inference from the gravitational wave data}
\label{appendix:LVK rate}

Let us now describe the inference process, used to obtain the GW coalescence rate $\mathcal{R}$, plotted e.g.~in Fig.~\ref{fig:example_rates}. The procedure we adopt is similar to that in ref.~\cite{Antoninietal2025b}. Concretely, we perform hierarchical population inference with a Hamiltonian Monte Carlo method, implemented in \texttt{numpyro} using \texttt{jax}. For a set of observed events $\set{d_i}$, the posterior on the population hyper-parameters $\Lambda$ is
\begin{equation}
    p(\Lambda \mid \{d_i\}) \propto p(\Lambda)\,\xi(\Lambda)^{-N_{\rm obs}}
    \prod_{i=1}^{N_{\rm obs}}
    \left\langle
    \frac{p(\theta_i \mid \Lambda)}{p_{\rm pe}(\theta_i)}
    \right\rangle_i \,,
    \label{eq:likelihood-sum-short}
\end{equation}
where $p_{\rm pe}(\theta_i)$ is the prior used in single-event parameter estimation, and the average is taken over posterior samples for event $i$~\cite[e.g.,][]{Fishbach_2018,2019MNRAS.486.1086M,2022ApJ...937L..13C}.
Selection effects are incorporated through the detection efficiency,
\begin{equation}
    \xi(\Lambda)=\frac{1}{N_{\rm inj}}
    \sum_{i=1}^{N_{\rm found}}
    \frac{p(\theta_i\mid\Lambda)}{p_{\rm inj}(\theta_i)},
    \label{eq:csi-short}
\end{equation}
computed by re-weighting recovered injections from the reference injection-distribution $p_{\rm inj}$~\cite{2022ApJ...937L..13C,2024PhRvD.109b2001A,injections,injection-methods}.

To control Monte Carlo noise, we monitor the effective number of posterior samples for each event,
\begin{equation}
    N_{{\rm eff},i}(\Lambda)=
    \frac{\left[\sum_j w_{i,j}(\Lambda)\right]^2}
    {\sum_j w_{i,j}^2(\Lambda)}\,,
    \qquad
    w_{i,j}(\Lambda)=\frac{p(\theta_{i,j}\mid\Lambda)}{p_{\rm pe}(\theta_{i,j})}\,,
    \label{eq:Neff-short}
\end{equation}
as well as the effective number of injections,
\begin{equation}
    N_{\rm eff}^{\rm inj}(\Lambda)=
    \frac{\left[\sum_i w_i(\Lambda)\right]^2}
    {\sum_i w_i^2(\Lambda)}\,.
\end{equation}
Following Ref.~\cite{2022arXiv220400461E}, we require $N_{\rm eff}^{\rm inj}\gtrsim 4N_{\rm obs}$.

Rather than imposing hard cuts, we suppress poorly sampled regions of parameter space by adding the penalty
\begin{equation}
    \ln S\!\left(\frac{N_{\rm eff}^{\rm inj}}{4N_{\rm obs}}\right)
    +
    \ln S\!\left(\frac{\mathcal{N}}{0.6}\right),
    \qquad
    S(x)=\frac{1}{1+x^{-30}},
\end{equation}
to the log-likelihood, where $\mathcal{N}\equiv \min_i \log N_{{\rm eff},i}$. This smoothly disfavours models for which either the injection set or one or more events are represented by too few effective samples.
We restrict our analysis to binary black-hole coalescence candidates in GWTC-4 with false-alarm rates below $1\,\mathrm{yr}^{-1}$, following Ref.~\cite{LVK2025GWTC4}. Data and posterior samples are taken from the publicly available LVK open-data releases~\cite{LVK_GWTC2.1_data_quality_2022,LVK_GWTC3_2023,LVK_GWTC4_2025_zenodo,LVK_open_data_2019,LVK_open_data_2023,LVK_open_data_2025}.

For events first reported in GWTC-1~\cite{LVK_GWTC1_2019}, we use the \emph{Overall posterior} parameter-estimation samples. For sources first presented in GWTC-2~\cite{LVK_GWTC_2}, we adopt the \emph{PrecessingSpinIMRHM} samples. For GWTC-3 events~\cite{LVK_GWTC3_2023}, we use the \emph{C01:Mixed} samples available on Zenodo, while for GWTC-4 events we use the \texttt{NRSur7dq4} samples whenever available~\cite{2019PhRvR...1c3015V}, and otherwise the \emph{Mixed} samples. We exclude systems with at least one component mass below $3\,M_\odot$, since such events are likely to involve a neutron star rather than constituting a binary black-hole merger~\cite{LVK_GWTC3_2023,LVK2025GWTC4,LVK_GWTC4_2025_zenodo}. After these selection cuts, the sample contains 153 events.
To account for observational selection biases, we use the publicly released set of successfully recovered injections produced by LVK for the first four observing runs~\cite{LVK_GWTC3_2023,injections,LVK2025GWTC4}. 

We assume that the merger-rate density can be expressed as
\begin{equation}
    \mathcal{R}(m_1,m_2,\chi_{\mathrm{eff}};z)
    =
    R_{\mathrm{ref}}\,
    \frac{f(m_1)}{f(20\,M_\odot)}
    \left(\frac{1+z}{1.2}\right)^{\kappa}
    p(m_2\mid m_1)\,p(\chi_{\mathrm{eff}}\mid m_1).
\end{equation}
where $m_2$ denotes the secondary mass, and $R_{\mathrm{ref}}$ is the merger rate per unit mass evaluated at $m_1 = 20\,M_\odot$ and $z = 0.2$. 

In the hierarchical analysis, we infer the joint distributions of the primary mass $m_1$, mass ratio $q$, redshift $z$, and the effective spin $\chi_{\rm eff}$. Following ref.~\cite{Antoninietal2025b}, we represent the primary-mass spectrum using a Gaussian process (GP),
\begin{equation}
    f(m_1) = \exp\!\bigl[\Phi(\ln m_1)\bigr],
    \qquad
    \Phi(x) \sim \mathrm{GP}\!\left(0,\,k(x,x';a_m,\ell_m)\right),
\end{equation}
where we adopt a squared-exponential kernel. The hyper-parameter $a_m$ sets the overall amplitude of the GP, controlling  vertical variation, while $\ell_m$ determines its correlation length and therefore the smoothness of the inferred spectrum. Both are treated as free hyper-parameters. The GP is evaluated on a uniform grid in $\ln m_1$ over the interval $2$--$200\,M_\odot$ and then interpolated to the event posterior samples and injection set.

The conditional distribution of the secondary mass is taken to follow
\begin{equation}
    p(m_2\mid m_1) \propto m_2^{\beta_q(m_1)},
    \qquad
    2\,M_\odot \le m_2 \le m_1.
\end{equation}
Here 
\begin{equation}
    \beta_q(m_1) = \Xi\!\left[\ln(m_1)\right],
    \label{eq:betaq_gp}
\end{equation}
where $\Xi \sim \mathrm{GP}\!\left(0,\,k(x,x';a_\beta,\ell_\beta)\right),$ is drawn from a GP prior defined on a regular grid in $\ln m_1$. As before, this GP is controlled by hyper-parameters governing its typical amplitude and smoothness. In particular, we adopt an exponential covariance kernel with variance $a_{\beta_q}$ and correlation length $\ell_{\beta_q}$.

In our analysis we adopt an effective spin model that transitions at $m_1=\tilde m$, for some threshold mass $\tilde{m}$ (designed to approximate $m_\ast$), from a Gaussian to a uniform distribution with independent bounds, respectively. We treat the bounds of the uniform component, $\chi_{\rm eff,min}$ and $\chi_{\rm eff,max}$, as free parameters inferred from the data:  
\begin{equation}
    \label{UL_ind}
    p(\chi_{\mathrm{eff}}\,|\,m_1) =
    \left[1-\eta(m_1)\right]\,\mathcal{N}(\chi_\mathrm{eff};\mu,\sigma)
    + \eta(m_1)\,\mathcal{U}(\chi_{\mathrm{eff}};\, \chi_{\rm eff,min}, \chi_{\rm eff,max}).
\end{equation}
We place uniform priors on these bounds: $\chi_{\rm eff,max}\sim \mathcal{U}(0.05,1)$ and $\chi_{\rm eff,min}\sim \mathcal{U}(-1,\chi_{\rm eff,max})$.
The mixing function is
\begin{equation}
    \eta(m_1) = \left[ 1+\frac{1}{9}\exp\!\left(-\frac{m_1-\tilde{m}}{M_\odot}\right) \right]^{-1}.
\end{equation}
By construction, this gives $\eta(\tilde{m})=0.9$, so that at $m_1=\tilde{m}$ the Gaussian component accounts for $10\%$ of the total mixture.

For the redshift evolution, we assume that the volumetric merger rate scales as a power-law in $(1+z)$~\cite{Fishbach_2018,Callister_2020}. Under this assumption, the redshift distribution is
\begin{equation}
    p(z) \propto \frac{1}{1+z}\,\frac{\mathrm{d}V_c} {\mathrm{d}z}\,(1+z)^{\kappa}.
\end{equation}

\begin{figure*}
    \centering
    \includegraphics[width=\textwidth]{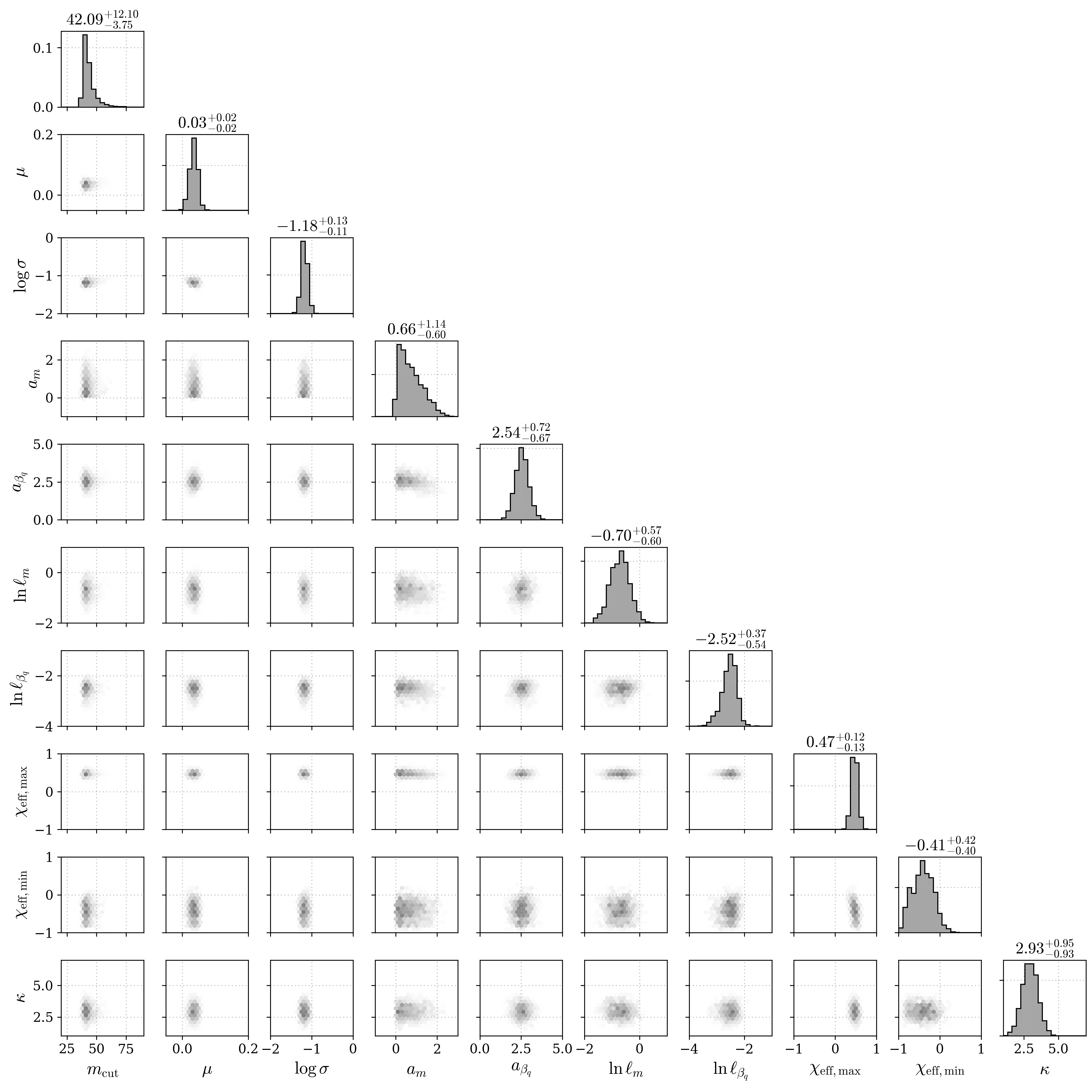}
    \caption{The posterior predictive distributions of the hyper-parameters of the population model.}
    \label{fig:post}
\end{figure*}

We give the priors adopted for all the hyper-parameters of the population model in table \ref{tab:priors}. Fig.~\ref{fig:post} displays the recovered posterior distributions.

\begin{table}
    \centering
    \renewcommand{\arraystretch}{1.05}
    \begin{tabular}{|c l|}
        \hline
        Parameter & Prior  \\
        \hline
        $a_{\beta}$
        & $\mathcal{HN}(0.7)$\\
        $\ln \ell_{\beta}$
        & $\mathcal{N}(0.5,0.35)$  \\
        \hline
        $a_{m}$ &$\mathcal{HN}(0.8)$\\
        $\ln \ell_{m}$  & $\mathcal{N}(-0.2,1)$   \\
        \hline
        $\tilde{m}$
        & $\mathcal{U}(20,100)$ \\
        $\mu$  
        & $\mathcal{U}(-1,1)$ \\
        $\sigma$   
        & $\mathcal{LU}(-1.5,0)$  
        \\
        $\kappa$ & $\mathcal{N}(0,6)$ \\
        $\chi_{\rm eff,\;max}$ & $\mathcal{U}(0.05,1)$          \\
        $\chi_{\rm eff,\;min}$ & $\mathcal{U}(-1,\chi_{\rm eff,\;max})$  \\
        \hline
    \end{tabular}
    \caption{
        Priors adopted for the hyper-parameters of the population model. 
    }
    \label{tab:priors}
\end{table}
\clearpage

\end{document}